# Recent advances on $CO_2$-assisted synthesis of metal nanoparticles for the upgrading of biomass-derived compounds

Zhiwei Jiang [a, 1], Yongjian Zeng [a, 1], Ruichao Guo [a], Lu Lin [a], Rafael Luque [c, d, **], Kai Yan [a, b, *]

[a] *School of Environmental Science and Engineering, Sun Yat-Sen University, Guangzhou 510006, China*

[b] *Guangdong Laboratory for Lingnan Modern Agriculture, South China Agricultural University, Guangzhou 510642, China*

[c] *Departamento de Química Orgánica, Universidad de Córdoba, Campus Universitario de Rabanales, Edificio Marie Curie (C3), Córdoba E-14014, Spain*

[d] *Department of Natural Sciences, Mid Sweden University, Holmgatan 10, 85170 Sundsvall, Sweden*

[*] Corresponding author.

[**] Corresponding author.

*E-mail addresses*: qo2balua@uco.es (R. Luque), yank9@mail.sysu.edu.cn (K. Yan).

[1] The authors contributed equally to this work.

**Abstract:** Nanostructured catalysts have attracted the increased attention for biomass conversion into high-valued chemicals due to the rapid depletion of fossil resources and increasingly severe environmental issues. Supercritical carbon dioxide ($scCO_2$) fluid is an attractive medium for synthesizing nanostructured materials due to its favorable properties. In this review, the properties of $scCO_2$ and the roles of $scCO_2$ in the fabrication of metal nanoparticles were assessed in detailed. A general overview of the synthesis of different types of metal nanoparticles (including metal oxide nanoparticles) using $scCO_2$ and the relationship between the structure of the obtained metal nanoparticles and the preparation conditions such as reaction temperature and pressure, types of metal precursors, and deposition time are system summarized and compared in tables. Besides, compared to the meatal catalysts using the conventional methods, the catalysts obtained using $scCO_2$ exhibited excellent catalytic performance on biomass conversion reactions, mainly focused on oxidation, hydrogenation reactions. Finally, opportunities and challenges of metal nanoparticle preparation using $scCO_2$ for biomass valorization to chemicals and liquid fuels are highlighted. This review could be helpful for the rational design of more efficient metal catalysts for the selective synthesis of fine chemicals and fuels from biomass-derived chemicals.

**Abbreviations:** $scCO_2$: supercritical carbon dioxide; NPs: nanoparticles; SCFD: supercritical fluid deposition; SCFs: supercritical fluid; $T_c$: critical temperature; $P_c$: critical pressure; GO: graphene oxide; $Pd(hfac)_2$: palladium hexafluoroacetylacetonate; RFA: resorcinol–formaldehyde aerogel; CFD: chemical fluid deposition; CNT: carbon nanotube; PET: poly(ethylene terephthalate); LA: levulinic acid; GVL: γ-valerolactone; TPA: terephthalic acid; BAL: benzyl alcohol; TBHP: tert-butyl hydroperoxide; PMS: peroxymonosulfate; MWCNTs: multiwalled carbon nanotubes; CVs: cyclic voltammograms.

## 1. Introduction

The rapid depletion of fossil resources and increasingly severe environmental pollution have prompted the wide-ranging search for alternative feedstocks and technologies to build a sustainable, renewable society [1-3]. Many countries have launched some green strategies to achieve this goal by developing renewable chemicals and clean production technologies. “2030 Agenda for Sustainable Development”, which aimed at achieving sustainable development goals, has been adopted by the United Nations in 2015 [4]. China strives to achieve peak $CO_2$ emissions before 2030 and carbon neutrality before 2060 [5]. Biomass has been deemed one promising renewable alternative to carbon-neutral energy that can be utilized as feedstocks to fabricate renewable fuels and value-added chemicals [6-8]. Recently, specific feedstocks such as carbohydrates [9], triglycerides [10], glycerol [11], 5-hydroxymethylfurfural [12], cellulose [13], hemicelluloses and pentoses [14], lignin [15], and lignocellulose [16] through chemical catalytic technology have been widely exploited as platform chemicals to product efficiently various chemicals and biofuels for different industrial applications. During the biomass conversion, the abundant C/O-H, C-C, and C-O bonds in biomass could be activated using multiple multifunctional catalysts through oxidation or reduction reaction to generate various fuels and value-added chemicals [17-19]. Among these catalysts, metal-based catalysts such as ruthenium (Ru), platinum (Pt), gold (Au), palladium (Pd), iridium (Ir), rhodium (Rh) , copper (Cu), nickel (Ni), and cobalt (Co) metals have been extensively applied for biomass catalytic conversion reaction (oxidation, hydrogenation, ammonization) [20-29]. Generally, metal-based catalysts nanoparticles (NPs) can be fabricated throng the deposition of metal NPs on different supports using different techniques, such as incipient wet impregnation[30], homogeneous deposition precipitation [31], chemical vapor deposition [32], photo-assisted depositions [33], and deposition-precipitation method [34].

Supercritical fluid deposition (SCFD) has become a bright, environmentally friendly method because of needless organic solvent, low viscosity, and high diffusivity, using which various metal and metal oxide materials with high dispersion and ultrafine particles have been produced. Well-defined nanoscale metal can be deposited on the solid support by reducing metal salts or organometallic precursors during the SCFD process. The hydrogen as a reducing agent allows the synthesis of pure metal nanoparticles with a green method without using organic solvent [35]. Moreover, thermal decomposition and chemical reduction under different atmospheres can also occur after deposition [36,37]. Several studies have revealed that metal particles with controllable properties such as size distribution, morphology, catalytic activity, porosity, density, and others could be generated by changing the conditions during the SCFD process [38-40]. Among supercritical fluid compounds, supercritical carbon dioxide ($scCO_2$) is a green solvent because of its nontoxic, renewable, and economical properties [41]. Furthermore, using $scCO_2$ to fabricate high-performance catalysts has been pinpointed.

Until recently, some reviews have systematically reported the development and synthesis of metal materials using $scCO_2$ for various applications [42,43]. However, summaries on the prepared metal materials through $scCO_2$ SCFD method for volatilizing biomass-derived compounds are rarely reported. This review focuses on the recent advance in $CO_2$-assisted preparation of metal catalysts for volatilizing biomass-derived compounds into fuels and high-value chemicals. We first discuss the properties of $scCO_2$ and the roles of $scCO_2$ in forming metal nanoparticles. Different types of metal nanoparticles (including metal oxides) using $scCO_2$ and the preparation process are discussed. Moreover, the catalytic systems using the metal catalysts through $scCO_2$ SCFD method for the conversion reaction of biomass-derived compounds, mainly focusing on oxidation and hydrogenation reactions, are summarized and commented. Finally, the existing issues and challenges of metal nanoparticle preparation using $scCO_2$ for biomass valorization are also highlighted.

## 2. Properties of $scCO_2$

A supercritical fluid (SCFs) is defined as a fluid above the critical point, at which the densities of one compound's liquid and gas phases become identical. As we all know, when the environmental temperature and pressure increase, the gas phase becomes denser, and the liquid phase becomes light but the density decreases, as shown in Fig. 1 [44]. The SCFs exhibit a combination of gas and liquids physical properties and become identical phases, leading to an intrinsic transport property. The viscosity and diffusivity of SCFs show a unique property with low viscosity, surface tension, and high diffusivity; however, its density value is similar to that of liquid.

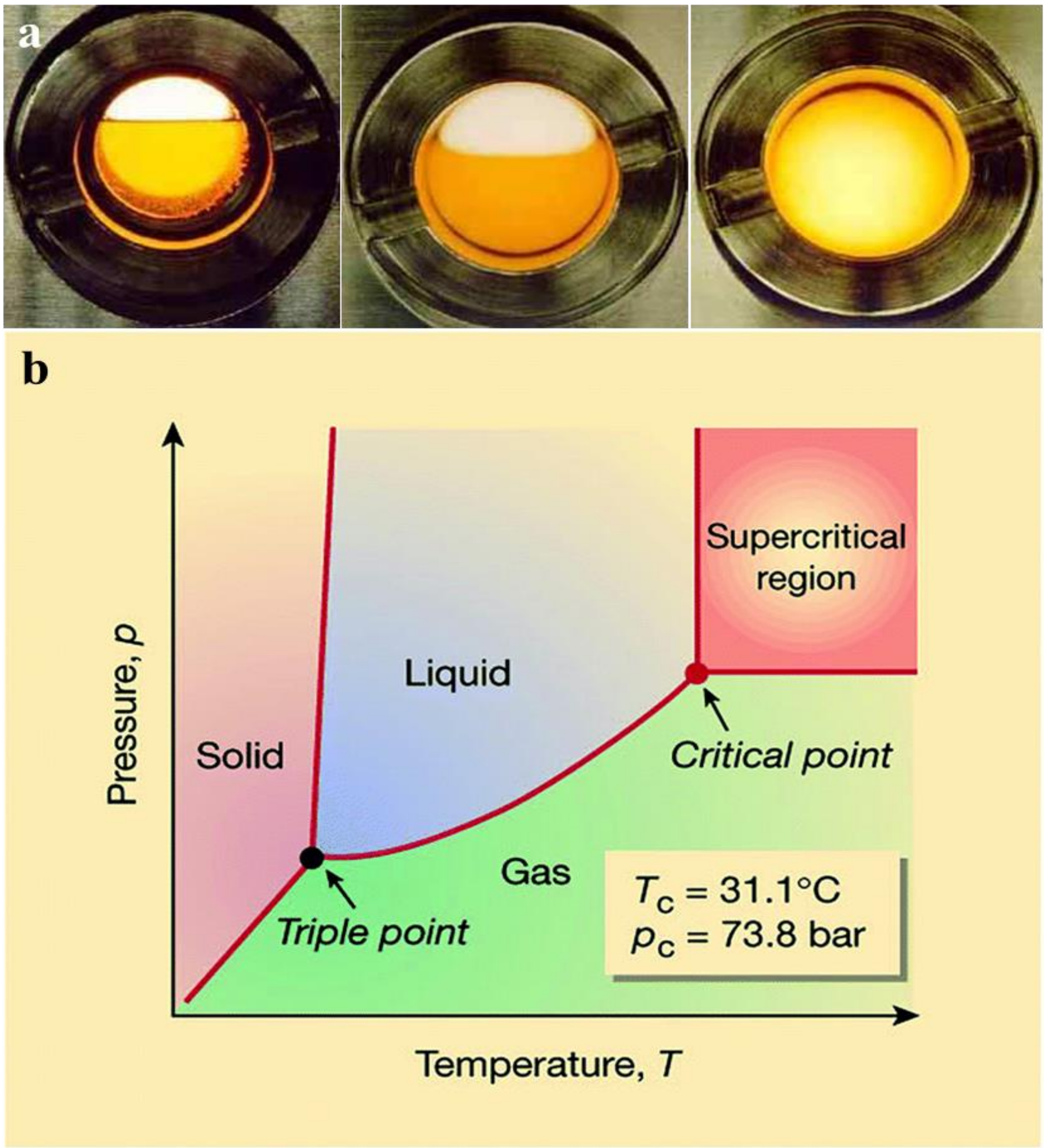

**Fig. 1.** (a) Phase transition pictures of coexisting gas-liquid (left) to the supercritical $CO_2$ phase (right); (b) Schematic representation of the pressure-temperature phase diagram ($T_c$: critical temperature, $P_c$: critical pressure). Reproduced from Ref. [44] with permission from Springer Nature, Copyright 2000.

For the fluid properties, SCFs have been applied in several fields, such as nanomaterials preparation, extraction, and drying [38,45], and have achieved plenty of attention in the biomedical, pharmaceutical, and food industries [46-48]. Besides, particle design and synthesis, extraction of chemical compounds, compound impregnation, and inactivation of microbes and enzymes were also carried out using SCFs [49-51].

**Table 1** Critical properties of different SCFs.

| Compound | $T_c$ (°C) | $P_c$ (MPa) | Remarks |
|---|---|---|---|
| Ammonia | 132.4 | 11.29 | Irritant and corrosive |
| Benzene | 289.1 | 4.9 | Carcinogen and high critical temperature |
| Carbon dioxide | 31 | 7.38 | Low critical temperature |
| Ethanol | 243.1 | 6.39 | High critical temperature |
| Isopropanol | 235.6 | 5.37 | High critical temperature and low toxicity |
| Methanol | 240.6 | 7.9 | High critical temperature |
| n-Propane | 93.9 | 4.3 | Asphyxiant |
| Propane | 96.8 | 4.26 | Asphyxiant |
| Water | 374.1 | 22.1 | Carcinogen properties and high critical temperature |

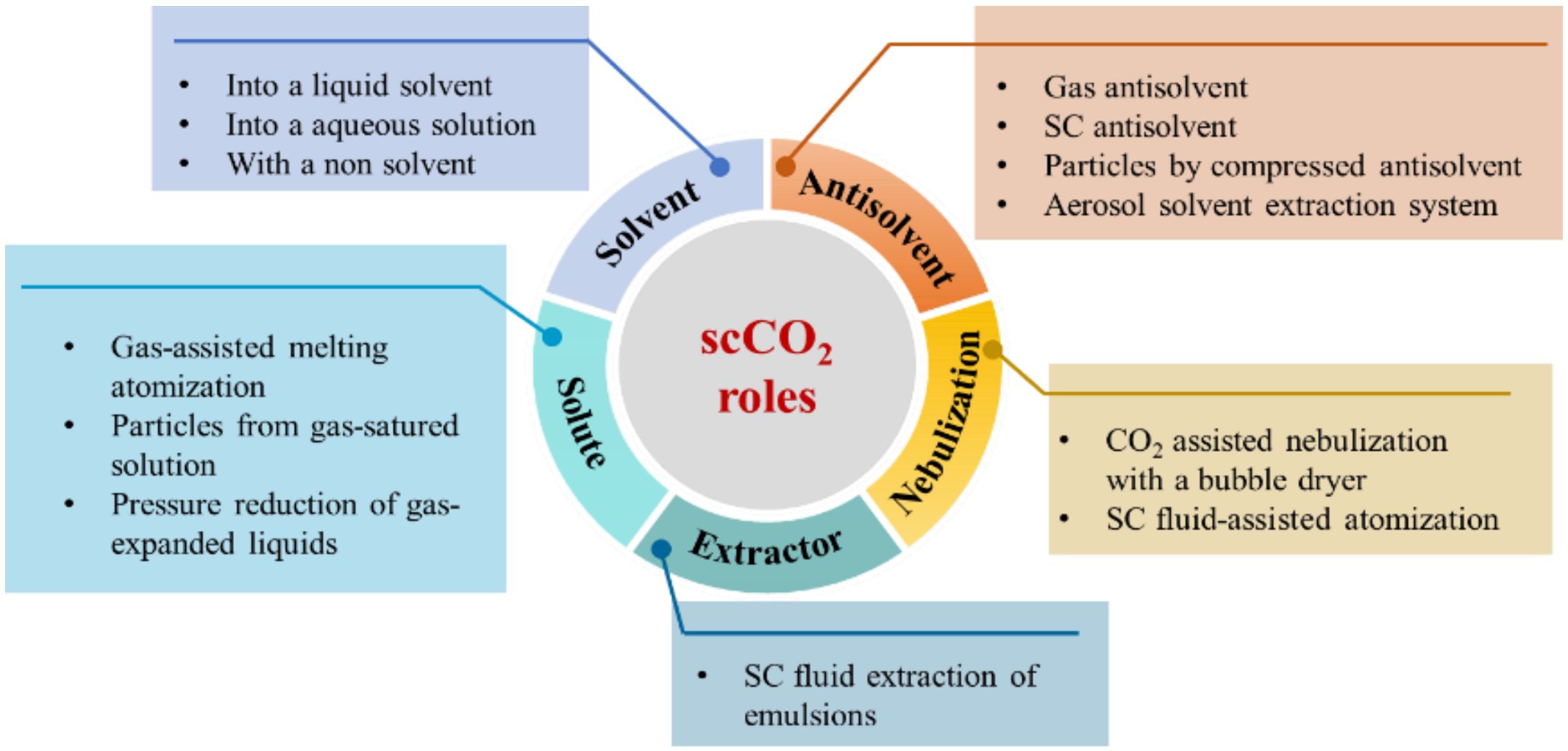

**Fig. 2.** Several $scCO_2$ roles on the particle formation. Reproduced from Ref. [43] with permission from Elsevier, Copyright 2021.

Table 1 shows several substances ($CO_2$, ethanol, isopropanol, water, *n*-propane and so on) that can be exploited to form SCFs, and their critical points determine the application field [52]. Water has been extensively utilized as SCFs because of its high availability, non-toxicity, and low cost. However, the high critical point of $H_2O$ ($T_c$ 374 °C and $P_c$ 22 MPa) limits its application fields. Thus, $scCO_2$ with relatively suitable critical pressure and temperature (31 °C and 7.4 MPa) is a promising SCFs and has gained considerable attention. Noted for presenting these advantages, the excellent solubility ability of $scCO_2$ allows its application in fabricating materials, resulting from the quick separation of products after depressurization. Moreover, due to its low critical temperature, $scCO_2$ has been exploited in non-thermal processing, in which thermosensitive compounds (drugs and phytochemicals) have been extracted. Besides critical points, other properties such as toxicity, corrosivity, and flammability should also be considered to choose suitable SCFs. It is essential to search for alternatives to avoid using toxic and hazardous solvents during material manufacturing due to environmental issues. As a nontoxic, inert and non-flammable solvent, $scCO_2$ has tremendous potential for various material production processes [53]. The production of metal particle materials can be achieved through $scCO_2$ usage. The role of $scCO_2$ can be classified in the material production process, as shown in Fig. 2. $ScCO_2$ can be used as an extractor, solute, solvent, anti-solvent, or solution nebulization [38].

## 3. $scCO_2$-assisted synthesis of catalytic metal nanoparticles

### *3.1. $scCO_2$-assisted synthesis of metal nanoparticles*

Using Supercritical technology to synthesize and design particles is an environmentally friendly method that provides a feasible route to sustainability. During this process, no contaminant residues are generated to avoid the pollution of the atmosphere, river, and soil [54]. It is worth noting that the interaction between metal particles and supports plays an essential role in its catalytic properties. Moreover, the manufacturing process of catalysts could significantly affect the interaction. It has been reported that synthesizing the catalyst through SCFs technology allows the as-prepared materials to have excellent catalytic performance in various chemical reactions [55].

The SCFs deposition is a saturation or penetration process during which a compound or particle deposits uniformly onto an appropriate support surface. The physical and chemical properties of support materials can be modified by depositing an active compound onto the surface or bulk of support materials. SCFD is a more effective deposition process because of low diffusion, homogenous impregnated overload and long contact time [56]. Sc$CO_2$ fluid deposition has been applied to generate materials because sc$CO_2$ fluid has some advantages such as inflammability, non-toxicity, greenness, low cost, and recovery during processing [36]. Moreover, adding co-solvent could improve the solubility of various precursors to enlarge the application range of sc$CO_2$ fluid deposition [57]. sc$CO_2$ fluid deposition allows high-performance catalytic applications to develop sustainable industrial processes in green chemistry engineering [58]. Besides, metal NPs, including Pt [59-61], Pd [45,62], Ru [22,63,64], Au [24,65] and silver (Ag) [66] show excellent catalytic activities for some essential chemical reactions such as oxidation, hydrogenation, and amination. Usually, the diameter of NPs with well-catalytical performance is about 2-10 nm. The particles have a low specific surface area and strong surface adsorption energy, which causes them to generate agglomerates easily. To inhibit the aggregation, NPs are usually deposited on porous substrates with high surface areas to improve their dispersion.

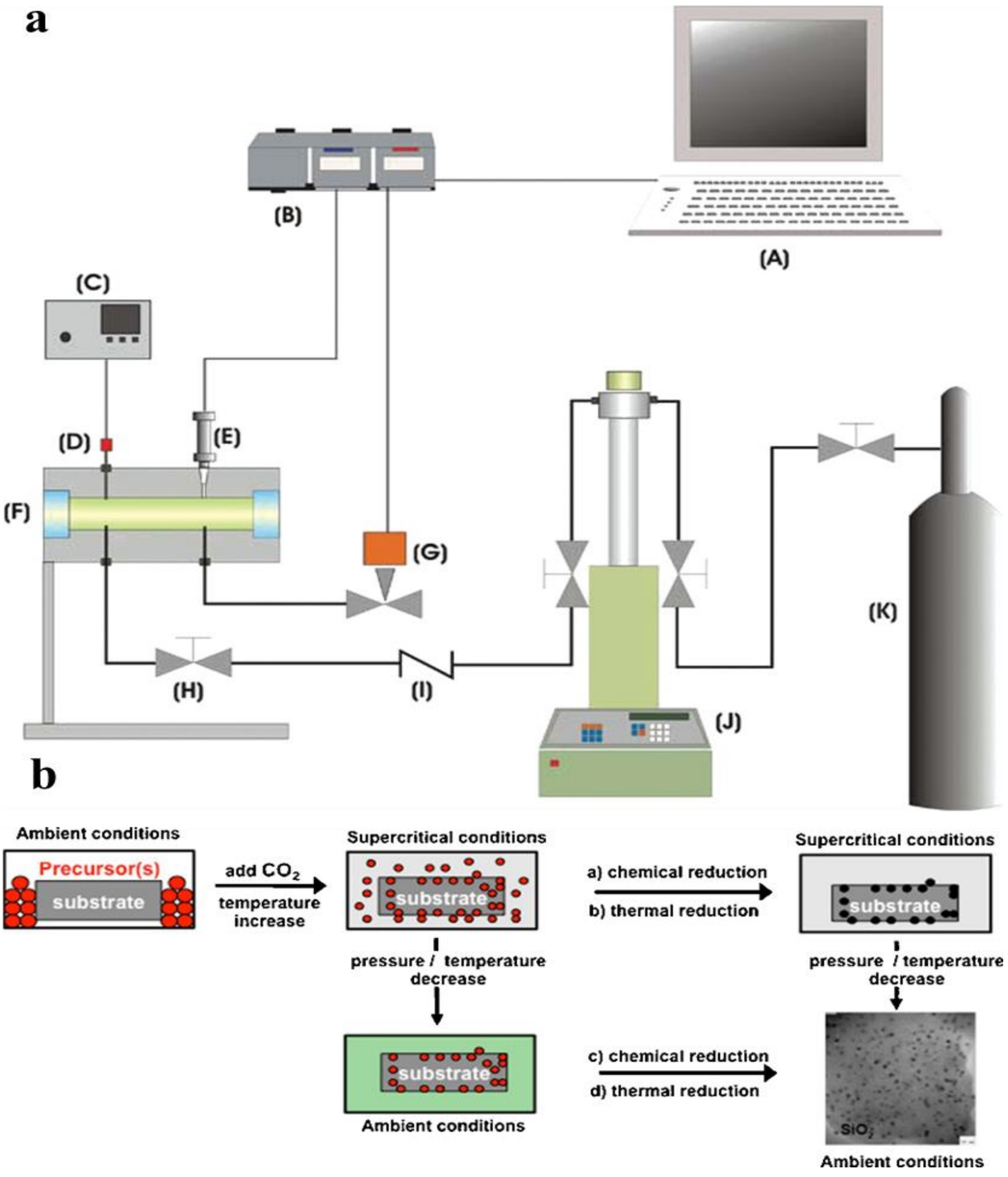

**Fig. 3.** (a) Schematic of a batch reactor for $scCO_2$. Reproduced from ref. [67] with permission from American Chemical Society, Copyright 2012. (b) Schematic representation of the formation of metal NPs. Reproduced from Ref. [54] with permission from Elsevier, Copyright 2018.

The basic principles of the formation of metal NPs through $scCO_2$ SCFD process can be illustrated in Fig. 3. Firstly, SCFs (mainly $scCO_2$) are used as a solvent to dissolve a metal precursor, followed by uniform deposition of the precursor onto appropriate support, and finally the formation of metal particles through the precursor reduction process. Several methods for the reduction formation of metal from the metallic precursor are as follows: a) chemical reduction using $H_2$ in $scCO_2$, b) thermal reduction at high temperature in $scCO_2$, c) chemical reduction with $H_2$ at low pressure, and d) thermal reduction at ambient pressure and high temperature in an inert atmosphere. Furthermore, the formation of binary NPs also could be formed based on the above methods [68].

Through changing preparation conditions, including reaction temperature and pressure, types of metal precursors, and deposition time, the structure and size of the formed metal catalysts using $scCO_2$ SCFD process were discussed. For Pt@γ-$Al_2O_3$ catalysts[69], inorganic Pt metal salt precursor and organic Pt metal salt precursor were used and the obtained Pt nanoparticles were compared. For inorganic metal precursor $Na_2PtCl_6$, 41.8% dispersion of Pt particles could be achieved, while only 25.6% Pt dispersion could be obtained using organic metal precursor Pt(cod)$Cl_2$. Furthermore, the smaller particle size with $1.5 \pm 0.4$ nm occurred from the precursor $Na_2PtCl_6$. The role of co-solvent factor also be considered for the metal catalysts using $scCO_2$. Using ethanol as a co-solvent, the high particle dispersion could be achieved because of the high solubility of inorganic precursors in ethanol [70].

For reaction temperature, the dispersion of Pt salt precursor increased the temperature from 35 to 40 °C due to a high diffusion coefficient of $scCO_2$ fluid at high temperatures [69]. As expected, the smallest Pt size was formed at 40 °C. For reaction temperature increased from 9 to 17 MPa, the Pt particle size firstly decrease, followed by an increase. When the deposition time was 3 h, smaller Pt particles were obtained. Furthermore, Pt can't be deposited on the support within a short time.

The Pd/$Al_2O_3$ catalysts were prepared and discussed at 15–24 MPa of $CO_2$ and 45–60 °C [71]. Furthermore, the catalytic activity of different catalysts prepared at different process conditions also was investigated. It is worth noting that the catalytic reaction mainly happened on the surface of the catalyst rather than its inside. The Pd catalysts achieved at high temperatures and pressures could reduce their hydrogenation catalytic activity because of Pd nanoparticle aggregation. Depositing the second metal Ag promoted the hydrogenation performance and reduced the surface area.

Some studies about the difference in catalytic performance of catalysts obtained via supercritical deposition method and traditional techniques were also explored. For example, the Pt@$TiO_2$ catalyst was fabricated in sc$CO_2$ condition at 300 °C and 20 MPa (S-Pt@$TiO_2$) for photocatalytic hydrogen production [72]. Compared with Pt@$TiO_2$ achieved via traditional techniques (T-Pt@$TiO_2$), S-Pt@$TiO_2$ catalyst shows better catalytic activity because of high surface area and pore volume. Moreover, SCFD method is more environment-friendly and reliable than traditional deposition methods, which are complex and expensive. The metal particle size of metal materials can be verified through varying pressure, temperature, and the concentration of precursor during sc$CO_2$ SCFD method. Using graphene oxide (GO) as support, several metals (Pt, Ru, Ni) were deposited under different process conditions to synthesize catalysts with different metal particle size. [73]. When the reduction condition with $H_2$/$N_2$ at 200 °C changed to $H_2/CO_2$ at 80 °C, the metal particle size increased from 4.6 to 5.4 nm, revealing that metal materials obtained by the SCFD method possessed a smaller metal size compared with that achieved by traditional techniques. Besides, the catalytic activity obtained by SCFD method and traditional technique was also investigated. For the hydrogenation reaction of limonene, Ru/rGO, obtained by SCFD method, exhibited better catalytic activity with 77% conversion within 60 min. However, the catalyst obtained by the traditional technique showed a 42% conversion.

**Table 2** Summary of metal-based catalysts with different supports obtained by $scCO_2$.

| Catalyst | Temperature (°C) | Pressure (MPa) | Contact time (h) | Reduction Temperature (°C) | Reduction time (h) | Reference |
|---|---|---|---|---|---|---|
| Ru/C | 80 | 8 | 4 | 350 | 3 | [74] |
| $PdAg/Al_2O_3$ | 45-60 | 15-24 | 0.33 | 250 | 2 | [71] |
| $Pt/TiO_2$ | 300 | 20 | 2 | NA | NA | [72] |
| Pt/GPE | 60, 80 | 10, 13.5 | 24 | 200, 400 | 4 | [73] |
| PtCu/SBA-15 | 60-80 | 13, 13.5 | 1-4 | 200, 400 | 4 | [75] |
| Pd/CNT | 50 | 18 | 5 | NA | NA | [76] |
| Pt-Ru-Ni/GPE | 50 | 16 | 4 | NA | NA | [77] |
| Pd/polymer | 40 | 25 | NA | 35 | 3 | [78] |
| Pt/Vulcan XC-72 | 50 | 13.2 | 12 | Room, 200 | 4 | [79] |
| $Pt/TiO_2$, $Pt/Al_2O_3$ | 80 | 15.5 | 20 | 25, 80 | 2 | [80] |
| Pt/GPE | 200 | 16 | NA | 300 | 1.5 | [81] |
| PtRu/GPE | 200 | 16 | 0.5 | NA | NA | [82] |
| RhPt/SBA-15 | 40 | 17.24 | 2 | 400 | 2 | [83] |
| $Pd/SiO_2$ | 80 | 17.2 | NA | 80 | 8 | [84] |
| Pt/GPE | 35 | 10.7 | NA | 80-800 | 4 | [85] |
| Au/γ-$Al_2O_3$ and AuAg/$TiO_2$ | 80 | 15.5 | NA | 80 | 2 | [86] |
| Pt-Cu/C | 35 | 10.7 | NA | 200 | 2 | [87] |
| Pt/PC and PdPt/PC | 40-70 | NA | 0.5 | NA | NA | [88] |
| PtIrCo/C | 82 | 20 | 48 | 600, 900 | 1 | [89] |
| $Pt_{40}Fe_{60}$/GPE | 60 | 15.2 | 2 | NA | NA | [90] |
| Pt/GPE | 70 | 24.5 | 6 | 200 | 4 | [91] |
| Pt/C | 80 | 27.6 | NA | 200-1000 | 4 | [92] |
| Pt/CNT | 200 | 10 | 1 | 200 | 0.5 | [93] |
| Ag/CNF | 36 | 17.2 | 24 | 180 | 2 | [94] |

Besides process conditions, supports were also another important factor for the formation and catalytic activity of metal-based catalysts. Using the sc$CO_2$ deposition method, many metals-based materials with different supports were produced and listed in Table 2. Among these supports, zeolite as support has been extensively investigated. Cabañas et al. [95] used sc$CO_2$ to deposit Pd into SBA-15. Palladium hexafluoroacetylacetonate [Pd(hfac)$_2$] was utilized as Pd precursor and deposited into SBA-15 support in sc$CO_2$ with 85 bar at 40 °C. Then, the Pd precursor was reduced with $H_2/CO_2$ at low temperatures or $H_2/N_2$ at high temperatures. The Pd nanoparticles uniformly dispersed into the support and the size of Pd particle increased with the Pd precursor concentration.

Polymers have also been used as supports to synthesize metal nanoparticles using the sc$CO_2$ SCFD method. The solubility capacity of metal precursor in sc$CO_2$ determines the amount of metal precursor adsorption on the polymer, thus affecting the size and distribution of the metal particles on polymer support. Besides, the polymer plasticization in $scCO_2$ also plays an essential role in the deposition of metal precursors in the polymer after $CO_2$ depressurization. When the metal precursor/polymer composite was prepared, metal nanoparticles can be produced from the metal precursor by the low-temperature reduction or high-temperature reduction to form organic-inorganic hybrid materials [96].

Moisan et al. [97] report a novel route ($scCO_2$ SCFD method) for the synthesis of organic-inorganic hybrid nanoparticles based on the reduction of a metal precursor using a $CO_2$-insoluble polymer. The polymer could stabilize the obtained metal nanoparticle with a small size and a high distribution through $H_2$ reduction. After depressurization, the functionalized metal nanoparticles are achieved in the form of dry powders with any solvents, which are easy to disperse in a suitable solvent. This approach has been applied to prepare several functional nanoparticles using different polymer-based supports. For example, Pd and Ag nanoparticles with 3−5 nm size have been achieve on hyperbranched polyamines in $scCO_2$, which can be functionalized with polysiloxane, perfluoroalkyl, perfluorooligoether, polyethylene glycol, or non-fluorinated alkyl moieties. Besides, Hasell et al. [98] report a one-pot successful fabrication of Ag–polymer nanocomposite in $scCO_2$. The polymer microparticle was slowly grown through the reversible addition-fragmentation chain transfer method in the presence of a monomer dissolved in $scCO_2$ and the Ag organometallic precursor was stabilized and thermally decomposed on the surface of the obtained polymer to form Ag-polymer nanocomposite.

The fabrication of Pt nanoparticles on resorcinol–formaldehyde aerogel (RFA)

polymer pellets using the SCFD method through a flowing atmospheric pressure $H_2$ reduction process at 200 °C has been reported [99]. Pt(cod)me$_2$ was used as Pt precursor and its kinetics adsorption on RFA spheres at different $CO_2$ pressures was also investigated through the determination of Pt(cod)me$_2$ concentration in sc$CO_2$. Moreover, the Pt nanoparticles size with other Pt(cod)me$_2$ concentration were also performed by TEM, as shown in Fig. 4. The results confirmed that the Pt nanoparticle size increases from 2.0 nm to 3.3 nm with Pt loading content increasing from 10 wt.% to 34 wt.% and the nanoparticle size achieves a plateau with 22 wt.% Pt loading content.

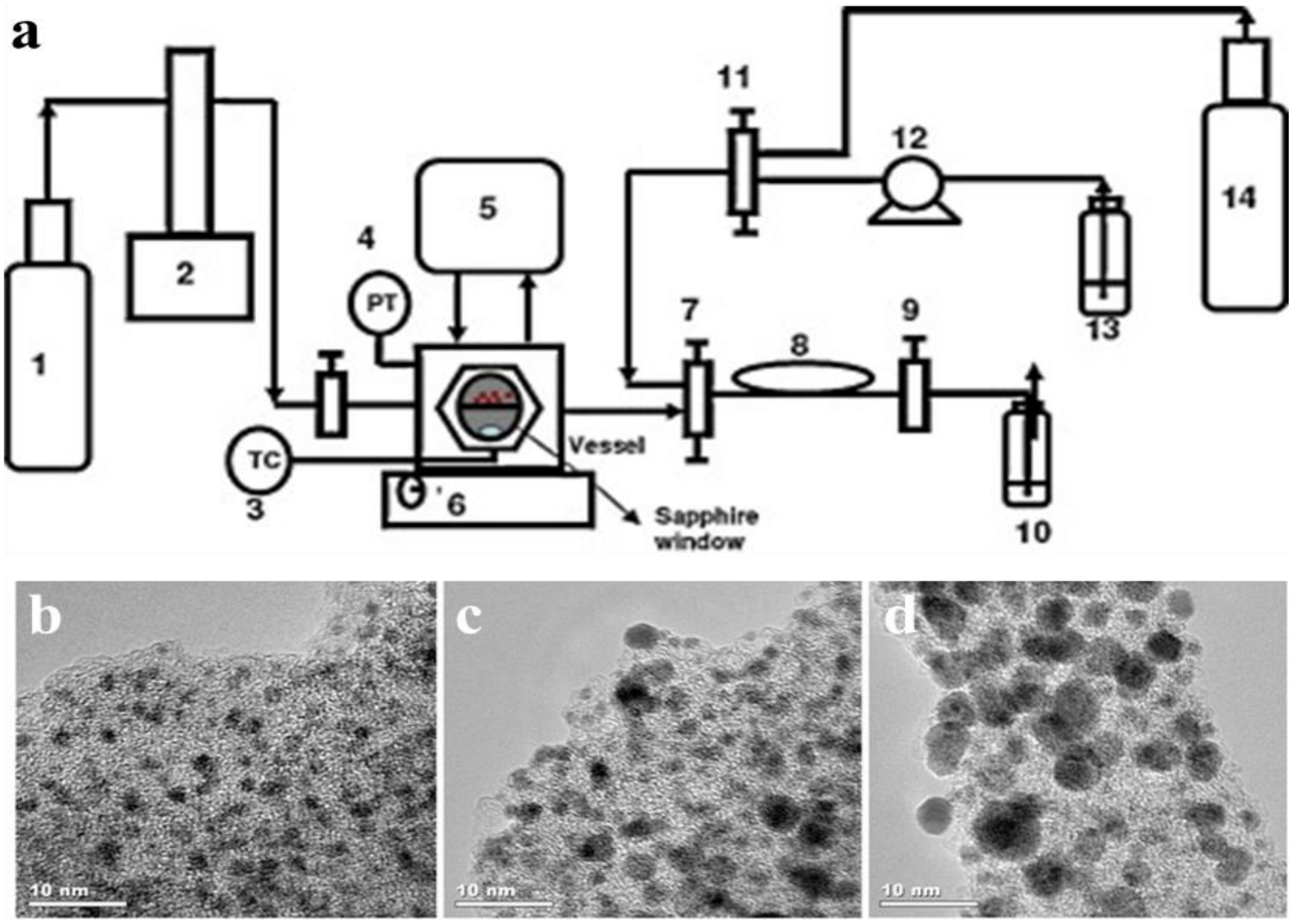

**Fig. 4.** (a) Diagram of the experimental setup. (b–d) BF-TEM images of Pt/RFA composites synthesized using SCFD with low-pressure $H_2$ conversion with Pt loading of 10, 22, and 34 wt.%, respectively. Reproduced from Ref. [99] with permission from Elsevier, Copyright 2011.

Besides mono-metal nanoparticles, bimetallic catalysts nanoparticles were synthesized through SCFD. Cabañas et al.[75] proposed the fabrication of Pt-based bimetallic catalysts using green $scCO_2$ as solvent. Pt, Cu and Ru metalorganic precursors deposited on SBA-15 were reduced to form PtCu and PtRu nanoparticles in $scCO_2$. The different reduction methodologies with different temperatures were used to deposit both metals simultaneously and sequentially. It is observed that the metal nanoparticles were uniformly deposited on the SBA-15 mesopores, confirmed by the TEM images (Fig. 5). Using $H_2/N_2$ at low pressure to reduce the metal precursor, smaller nanoparticles were received.

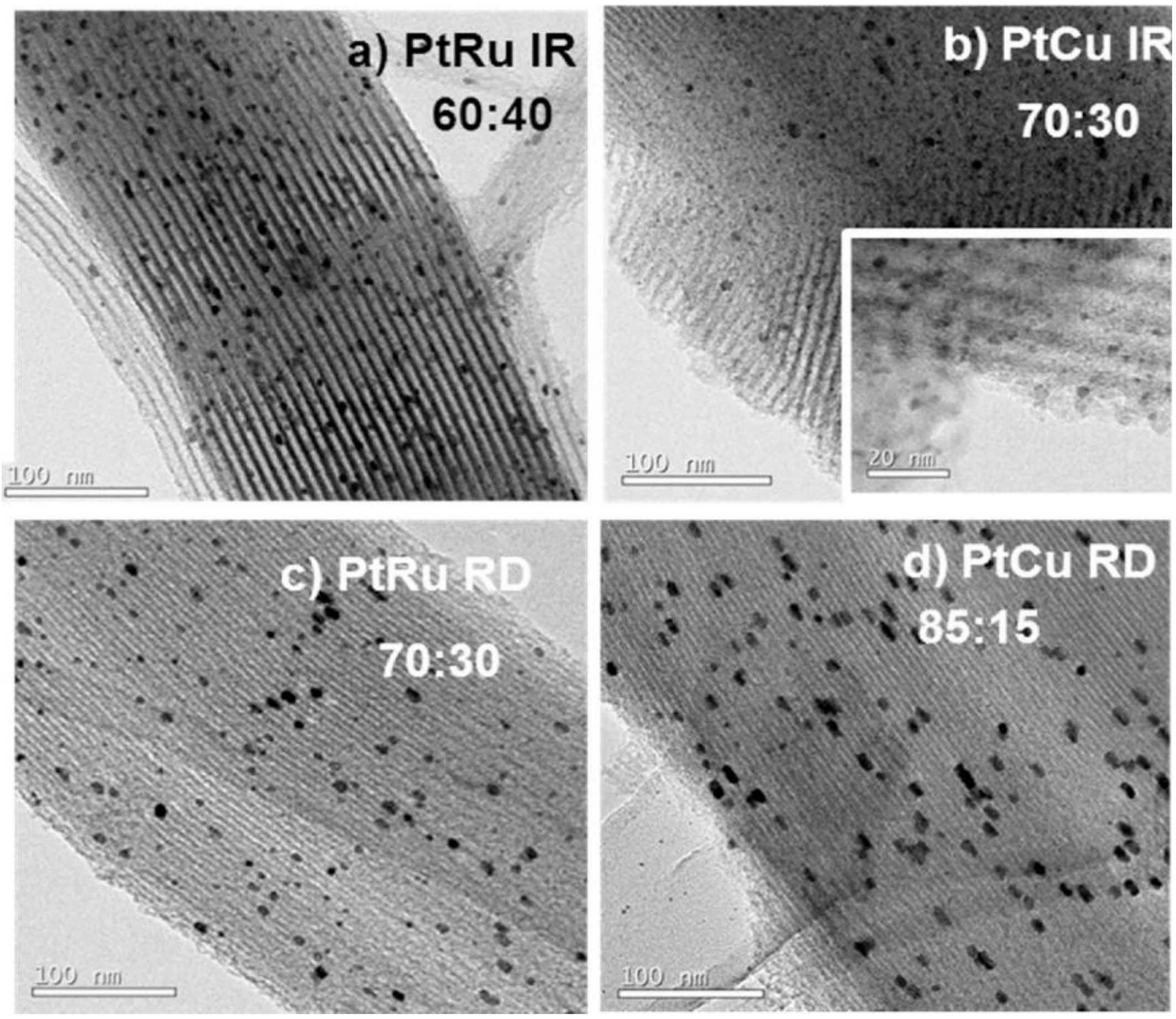

**Fig. 5.** TEM images and Pt to Ru or Pt to Cu molar ratios for the bimetallic samples on SBA-15 prepared by SCFD simultaneous deposition: (a) PtRu IR, (b) PtCu IR (c) PtRu RD and (d) PtCu RD (IR: impregnation/reduction, RD: reactive deposition).

Reproduced from Ref. [75] with permission from Elsevier, Copyright 2017.

Leitner et al. [45] reported the preparation of PtPd bimetallic nanoparticles on mesoporous SBA-15 support using $scCO_2$ as the reaction medium. The TEM results showed that highly well-distributed nanoparticles with an average 5 nm were absorbed preferentially on the surface of inside pores in SBA-15. In comparison with the conventional impregnation method using various liquid solvents, the obtained metal nanoparticle size using $scCO_2$ SCFD method was similar to that using n-pentane, but showed a significant difference in that using tetrahydrofuran or toluene.

Non-novel metals nanoparticles were also produced through SCFD. Meng et al. [100] synthesized Ni metal-deposited graphene oxide materials (Ni/GO) with the assistance of $scCO_2$ (Fig. 6). The obtained Ni nanoparticles with about 5 nm were well-distributed deposited on the surfaces of GO nanosheets. The as-prepared Ni/GO materials were applied in paraffin oil as lubricating additives and showed good tribological properties, which was confirmed through a four-ball tribometer to measure. Compared with pure oil, paraffin oil contained 0.08 wt % Ni/GO and showed a smaller friction coefficient (32% of pure oil) and wear scar diameter (42% of pure oil). Moreover, nano-Ni, GO nanosheets were all added to paraffin oil to perform their tribological properties, finding that the Ni/GO nanosheets exhibited optimum lubricating performance.

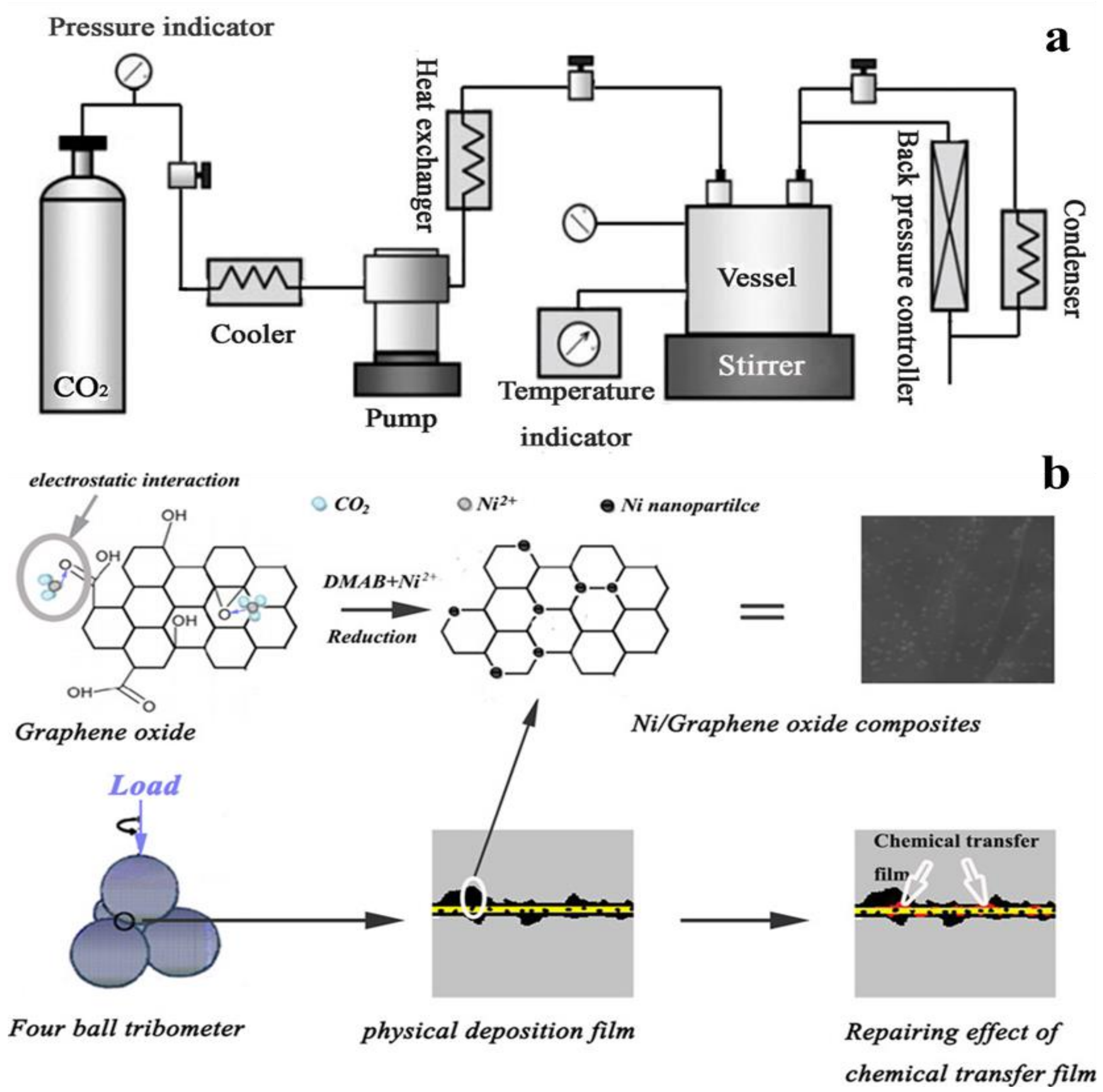

**Fig. 6.** (a) Diagram of the experimental setup. (b) The formation process of Ni/GO. Reproduced from Ref. [100] with permission from American Chemical Society, Copyright 2015.

Aymonier et al. [101] reported non-novel Mg-based nanoalloys using a modified SCFD approach. Cu and Ni nanoparticles were deposited on Mg metal support to synthesize $MgCu_2$ and $Mg_2Ni$ alloys under the mixed solution of isopropanol and $scCO_2$. The reduction process was performed using $N_2/H_2$ at 130 °C. The alloy formation was confirmed and at the interface was characterized by the TEM observations, revealing that the alloy formed at the interface of Cu or Ni metal nanoparticle and the Mg support and the crystal spacing of the alloys belong to $MgCu_2$ and $Mg_2Ni$ alloys.

Yan's groups [88] reported the formation of metal nanoparticle catalysts using the green liquid $CO_2$ fluids deposition method. The porous carbon derived from sorghum was used as support. Pd/C, Pt/C, and PdPt/C catalysts were successfully prepared at 40-70 °C with $N_2/H_2$. Moreover, the as-obtained catalysts exhibited excellent catalytic performance in the selectivity formation of furfuryl alcohol and tetrahydrofurfuryl alcohol from biomass-based furfural via hydrogenation reaction. Besides, Pd-based nanoparticles with various supports ($SiO_2$, $Al_2O_3$, MCM-41, ZrMCM-41, SnMCM-41, TiMCM-41) by chemical fluid deposition (CFD) methods were also fabricated, and their catalytic performance in the hydrogenation of furfural were compared [102]. Later, 5 wt.% Pd/AlMCM-41 were also successfully produced via a green liquid $CO_2$ fluids-assisted method for levulinic acid conversion into γ-valerolactone and valeric acid [103].

*3.2. $scCO_2$-assisted synthesis of metal oxide nanoparticles*

Metal oxide nanoparticles deposited on support were also prepared through $scCO_2$. Solubility of metallic precursors in $CO_2$ increases with the $CO_2$ densities, suggesting high solubility could occur at liquid or $scCO_2$. Peng et al. [104] prepared metal oxide nanoparticles deposited on silica and carbon nanotube substrates from metallic precursors exploiting the solvent properties of $scCO_2$. Metal oxide nanoparticles onto polymer matrices such as Nafion membranes also were facilitated using $scCO_2$ SCFD. [105] Zefirov et al. proposed the production of $MnO_x$ nanoparticles in $scCO_2$ from its metallic precursor. The as-obtained $MnO_x$ nanoparticle materials exhibited a smaller diameter of 42 nm, bigger specific surface area, and higher homogeneity, compared with that obtained using the conventional impregnation method under similar pyrolysis conditions. Moreover, the shape of the $MnO_x$ nanoparticles obtained via $scCO_2$ is rod-like rather than spherical [106]. The results can be explained that $scCO_2$ induced a "delayed precipitation" effect, and $scCO_2$ could improve the stability of the metallic precursor around the supports to prevent the aggregation and clustering of metallic precursor [106]. This suggests that $scCO_2$ can be used in the controllable synthesis of metal oxide nanoparticles with different sizes and shapes derived from metallic precursors that can be controllably fabricated by manipulating $scCO_2$ density and temperature.

$CoO_x$ nanoparticles anchored on porous silica support were also synthesized through $scCO_2$ supercritical fluid [107]. Firstly, Co metallic precursor was deposited on the porous silica at 70 °C with 11 MPa $scCO_2$ supercritical fluid for 3 h. Besides, the effect of different pressures ranging from 6 to 20 MPa on the content of Co metallic precursor on porous silica were carried out, demonstrating that the maximum adsorption content was 2.9 wt.% with 14 MPa. It is worth noting that the formed metal content increased with the thermal reduction time using $H_2$. The maximum Co content with 3.4 wt.% Co was obtained for 120 min during the thermal reduction of the Co precursor. The thermal reduction time also played an important role in the metal nanoparticles size.

Nickel oxide (NiO) nanoparticles deposited in carbon nanotube (CNT) arrays with millimeter thickness (NiO/CNT) were successfully produced using $scCO_2$ SCFD method [108]. The maximum content of NiO in NiO/CNT can be as high as 43.4 wt.%. The TEM images showed that carbon, nickel, and oxygen elements are dispersed homogeneously on NiO/CNT, suggesting that NiO nanoparticles were uniformly distributed on the CNT support. This can be explained that $scCO_2$ possessed high penetration ability, gas-like diffusivity and low surface tension, causing the Ni metallic precursors to be rapidly delivered and deposited uniformly on the CNT support, followed by allowing the distribution uniform of NiO nanoparticles in the support. Besides, $ZrO_2$/CNT were also synthesized using $Zr(NO_3)_4 \cdot 5H_2O$ precursor and CNT support through $scCO_2$ deposition method. The TEM images illustrated that a $ZrO_2$ layer was formed and deposited on the outer surface of CNT. The thickness of $ZrO_2$ layer can be regulated by changing the deposition time, the content of Zr precursors, and the type of Zr precursors. However, the $ZrO_2$ nanoparticles on CNT were synthesized using ethanol as solvent. The results confirmed that the formation of $ZrO_2$ layer coating on the CNT was ascribed to the low surface tension of $scCO_2$ and good wetting of CNT [109]. $SnO_2$ nanoparticles deposited CNT were successfully fabricated based on a similar $scCO_2$ process [110]. $SnCl_2$ was utilized as Sn precursor and dissolved in ethanol to form a homogeneous solution. Then, the mixture solution was added into a high-pressure vessel with $scCO_2$ and stirred vigorously for 5 h. After depressurization, $SnO_2$/CNT composite was achieved and then washed with ethanol and dried. $SnO_2$ nanoparticles were successfully anchored onto the surface of carbon nanotubes, based on the TEM images. The solvent of Sn precursor and the stirring time played significant roles in the size of $SnO_2$ nanoparticles.

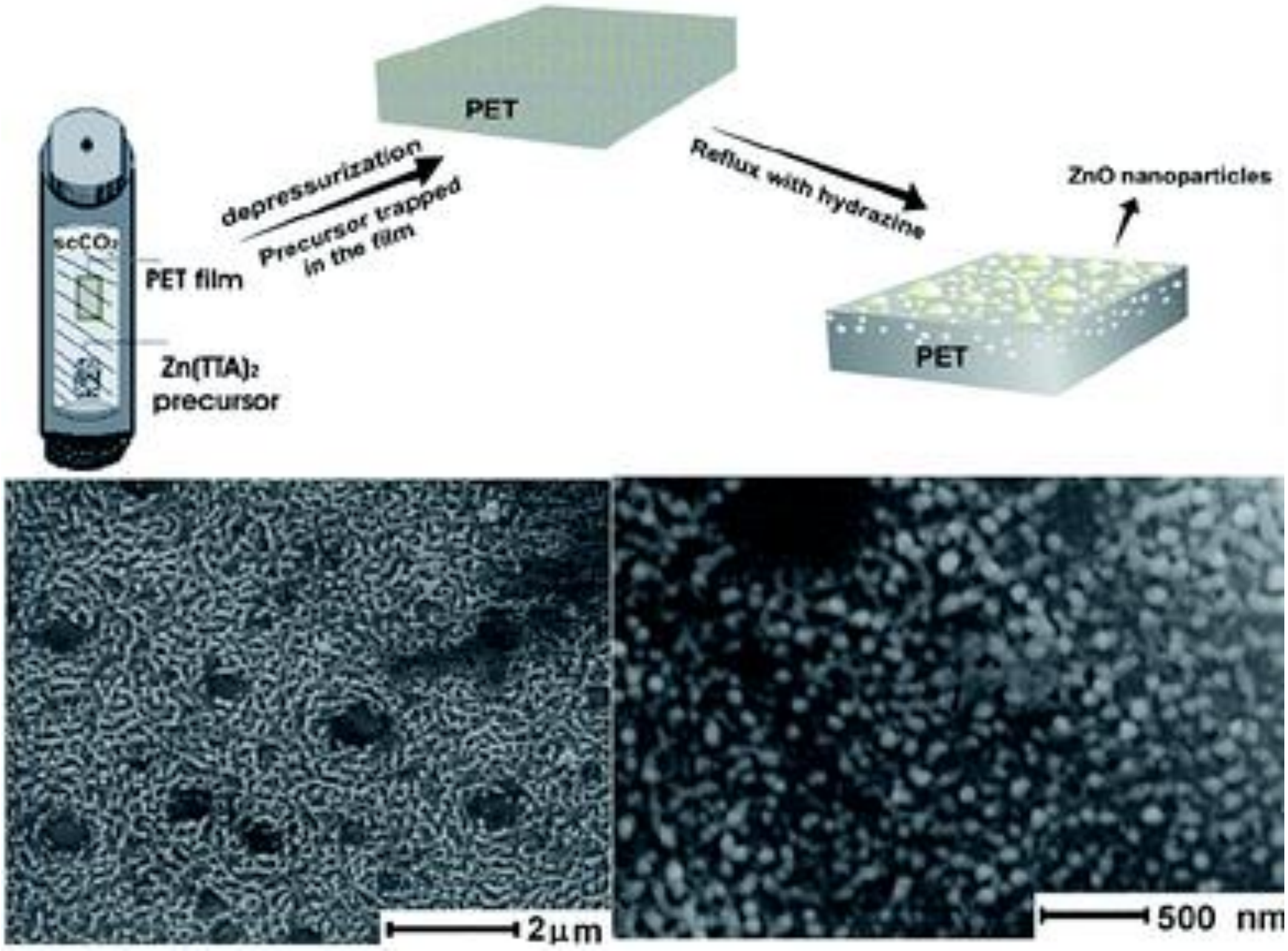

**Fig. 7.** Scheme representation of the deposition process and the SEM images of the resulting ZnO/PET composite. Reproduced from Ref. [111] with permission from Elsevier, Copyright 2011.

Recently, several metal oxides-polymer composites were reported by deposition with $scCO_2$ with tunable shape and catalytic properties. Mauricio et al. reported that ZnO nanoparticles were uniformly dispersed on poly(ethylene terephthalate) (PET) films to form ZnO/PET composite with the aid of $scCO_2$. Firstly, the zinc precursor was dissolved in $scCO_2$ and then slowly deposited onto the PET films, followed by treatment using $N_2H_4$-alcohol solution to form ZnO nanoparticles [111]. The formation process of ZnO/PET composite using $scCO_2$ and the SEM images of the obtained ZnO/PET composite are illustrated in Fig. 7. The effect factors on the Zn loading content were also investigated and achieved a maximum Zn loading content with 28% on the PET film. Moreover, the size of ZnO nanoparticles anchored on the PET film was 50–80 nm based on the SEM result.

Single atoms catalysts were also prepared through $scCO_2$ deposition. Qi et al. [112] reported that Cu single atoms were highly distributed on N-doped carbon nanosheets (Cu-NC) by $scCO_2$ deposition and decompression methods. Cupric (II) acetylacetonate was used as a precursor and the optimal process conditions are 50 °C and 20 MPa for 12 h stirring. The loading of Cu was 0.80% and the surface area of the as-prepared Cu-NC was as high as 1301 $m^2/g$. Besides, Cu-NC is beneficial for the formation of benzaldehyde from benzyl alcohol oxidation with excellent performance.

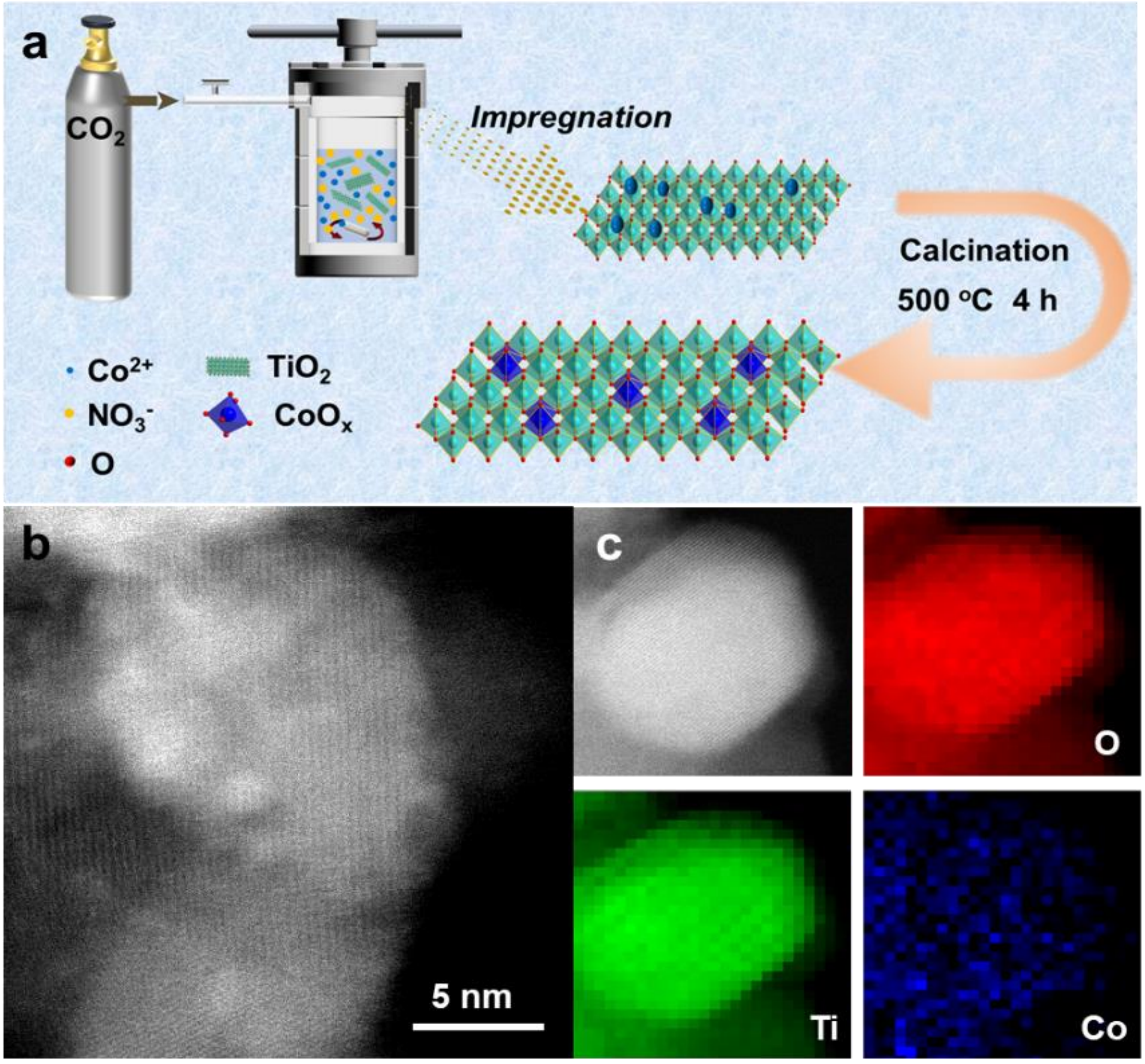

**Fig. 8.** (a) The scheme of the Co-$TiO_2$ formation, (b) HAADF-STEM image of $CoO_x$

clusters in Co-$TiO_2$, (c) STEM-EDS elemental maps for O, Ti, and Co, respectively. Reproduced from Ref. [113] with permission from Elsevier, Copyright 2023.

Besides sc$CO_2$, liquid $CO_2$ fluid was also applied to synthesize metal oxide nanoparticles deposited on supports. Jiang et al. reported that amorphous $CoO_x$ nanoparticles anchored in $TiO_2$ nanosheets (Co-$TiO_2$) with different Co content were prepared using the green liquid $CO_2$ solvent (Fig. 8) [113]. From the TEM images, amorphous $CoO_x$ nanoparticles with sizes of 0.7 to 1.5 nm were uniformly distributed on the $TiO_2$ nanosheet.

## 4. Catalytic conversion of biomass-derived chemicals

### *4.1. Hydrogenation*

The hydrogenation of C=O double bands (aldehyde and carbonyl groups) and other unsatisfied chemical bonds in biomass-based chemicals is a critical process to yield high-valued products. Metal nanoparticles catalysts exhibited excellent hydrogenation activity due to their superb ability for the activation of $H_2$. More importantly, the green method of $CO_2$-assisted preparation of metal nanoparticles has been widely applied in the hydrogenation of biomass-derived chemicals with commendable catalytic performance. The catalysts reported in the literature were summarized in Table 3.

**Table 3** Summary of catalysts prepared via sc$CO_2$ deposition used in the hydrogenation of biomass-based chemicals.

| Reactant | Catalyst and reaction conditions | Catalytic performance | Reference |
| --- | --- | --- | --- |
| Levulinic acid | $Pt_1Pd_3$/SBA-15, 220 °C, 2 h, 100 bar $H_2$ | 80% conversion, 80% γ-valerolactone | [45] |
| Levulinic acid | Pd/AlMCM-41, 240 °C, 10 h, 60 bar $H_2$ | 99% conversion, 88.5% γ-valerolactone | [103] |
| Levulinic acid | $Pd/Al_2O_3$, 160 °C, 6 h, 45 bar $H_2$ | 63.2% conversion, 60.9% γ-valerolactone | [114] |
| Furfural | RuPt(s)/SBA-15, 80 °C, 3 h, 18 MPa mixture of $CO_2$ and $H_2$ | 69% conversion, 28.0% furfuryl alcohol, 17.9% tetrahydrofurfuryl alcohol | [75] |
| Furfural | PdPt (2:1)/C, 140 °C, 3 h, 60 bar $H_2$ | 95% conversion, 70.3% furfuryl alcohol | [88] |
| Cinnamaldehyde | Rh/MSU-H, 50 °C, 3 h, 10 MPa $CO_2$, 4 MPa $H_2$ | 47.1% conversion, 40.0% hydrocinnamaldehyde | [115] |
| Benzene | Ru@Zr-MOF, 40 °C, 2 MPa $H_2$, 42 min | 99.5% cyclohexane | [116] |
| Terephthalic acid | $Rh_{70}Pt_{30}$/SBA-15, 80 °C, 4 h, 5 MPa $H_2$ | 99.6% conversion, 35.9% *cis*-1,4-CHDA, 63.7% *trans*-1,4-CHDA | [83] |
| Dimethyl oxalate | CuO/SBA-15, 200 °C, 3 MPa $H_2$, 1.2 $h^{-1}$ | 100% conversion, 90.0% Ethylene glycol | [117] |

Levulinic acid (LA) is considered to be a biomass-based platform molecule with wide usage because it can be converted into various high-quality products, for instance, γ-valerolactone (GVL) and valeric acid. $scCO_2$-assisted preparation catalysts have been widely used in the hydrogenation of LA. Leitner et al. [45] used mesoporous SBA-15 to support Pt/Pd bimetallic nanoparticles via a $scCO_2$ deposition method. The prepared catalysts were applied in the LA hydrogenation into GVL (Fig. 9a). The effect of Pt/Pd ratio on catalytic activity was studied (Fig. 9b). Besides, the traditional impregnation with different organic solvents was also used for comparison to investigate the influence of the $scCO_2$ medium. The nanoparticles of 3% $Pt_1Pd_3$/SBA-15 fabricated in $scCO_2$ exhibited a uniform distribution on SBA-15 support. The catalyst prepared in *n*-pentane had a similar morphology to the sample obtained by $scCO_2$ deposition, while the nanoparticles of the catalysts prepared in toluene and tetrahydrofuran agglomerated. The physico-chemical properties of $scCO_2$ and *n*-pentane were helpful to the mass transfer and deposition of metal precursors. Fig. 9c shows the catalytic performance of 3% $Pt_1Pd_3$/SBA-15 deposited in different mediums for LA hydrogenation. The catalytic performance was also affected by the deposition medium. 3% $Pt_1Pd_3$/SBA-15 fabricated in toluene and tetrahydrofuran (THF) only obtained ~30% yield of GVL. Obviously, 3% $Pt_1Pd_3$/SBA-15 fabricated in $scCO_2$ and *n*-pentane showed superior catalytic activity, reaching ~80% yield of GVL. Besides, the catalytic activity of the two catalysts had no obvious difference after 2 h reaction time (Fig. 9d), indicating that the properties of the deposition medium were important. The catalytic activity of 3% $Pt_1Pd_3$/SBA-15 deposited in $scCO_2$ only showed a slight decrease after the first used. There was no decrease during the following three times used (Fig. 9e). These results revealed that $scCO_2$ deposition played a suitable strategy for the preparation of catalysts, promoting the metal nanoparticles distributed on the pores of SBA-15 uniformly and resulting in the enhanced catalytic activity. Yan's groups also prepared [103] Pd nanoparticles supported on MCM-41 catalysts synthesized by the $CO_2$-assited method used for the hydrogenation of LA. The results showed that the acid sites on the support were crucial for the dehydration of the reaction intermediate of hydroxyvaleric acid. The prepared 5 wt.% Pd/AlMCM-41 catalyst with high acidity showed the highest catalytic activity for LA hydrogenation, achieving 45.1% yield of valeric acid at 270 °C. The mesoporous channel of MCM-41 could reduce the Pd nanoparticles aggregation efficiently.

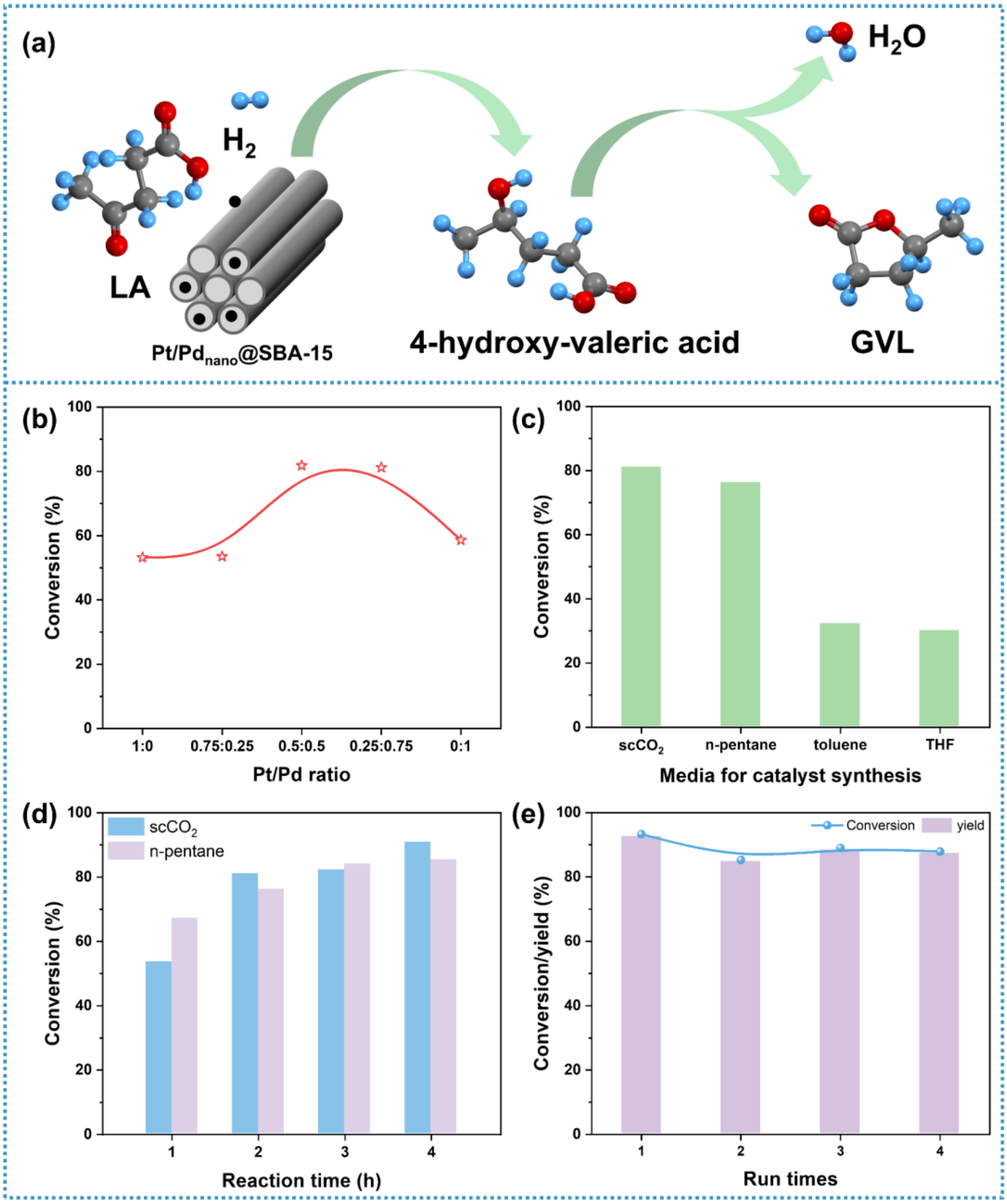

**Fig. 9.** (a) The reaction pathway of LA hydrogenation over Pt/Pd/SBA-15 catalyst. (b) Effect of Pt/Pd ratio on catalytic activity. (c) Catalytic performance of 3% $Pt_1Pd_3$/SBA-15 deposited in a different medium for LA hydrogenation. (d) Effects of reaction time on the catalytic performance of 3% $Pt_1Pd_3$/SBA-15 deposited in $scCO_2$ and n-pentane. (e) Recycle test of 3% $Pt_1Pd_3$/SBA-15 deposited in $scCO_2$. Reproduced from Ref. [45] with permission from Royal Society of Chemistry, Copyright 2017.

Furfural is another crucial platform chemical for the production of biofuels. Yan et al. [114] synthesized γ-$Al_2O_3$ supported highly dispersed Pd nanoparticle catalysts through the $scCO_2$ assistance for furfural hydrogenation. 39.4% conversion of furfural to furfuryl alcohol with 84.9% selectivity was obtained by 5 wt.% Pd/$Al_2O_3$ at 150 °C in 45 bar $H_2$. The d-band center of Pd metal could receive the electron transmitted from the α-bond of hydrogen, which promoted the hydrogenation reaction. In addition, the catalytic performance of 5 wt.% Pd/$Al_2O_3$ prepared via $scCO_2$ assistance was better than that of the catalyst synthesized by a traditional impregnation method, indicating that the reported $CO_2$ assistance method was superior. Cabañas et al. [75] used PtPt/SBA-15 catalysts via $scCO_2$-assisted preparation for the hydrogenation of furfural in the reaction medium of $scCO_2$. Pt/SBA-15 showed 73% conversion of furfural and 35% selectivity of furfuryl alcohol at 3 h. For Pt bimetallic catalysts, the order of deposition for the Ru metal particles could affect the catalytic activity. RuPt(s)/SBA-15 showed 70% conversion of furfural, higher than that of PtRu(s)/SBA-15 (40%). Although the conversion of furfural for Pt bimetallic catalysts was lower, the selectivity of furfuryl alcohol was higher than Pt catalyst. The lower catalytic activity of PtRu bimetallic catalysts may be caused by the lower Ru loading and larger particle size than RuPt catalyst. Yan's groups also investigated [88] Pd-Pt bimetallic catalysts on porous carbon support derived from sorghum biomass waste via $CO_2$ deposition strategy for furfural hydrogenation (Fig. 10a). The metal nanoparticles of all the prepared catalysts were dispersed on carbon support uniformly with no aggregation. The catalytic activity of different catalysts for furfural hydrogenation is shown in Fig. 10b. 3 wt.% Pd/C catalyst with $CO_2$-assisted preparation exhibited better catalytic activity than the commercial Pd/C for the furfural hydrogenation. The yield on the latter was 7 times of the former. The results showed that the carbonyl group and furan ring of furfural were preferentially hydrogenated by Pd/C and Pt/C catalysts, respectively. The products distribution could be adjusted by changing the radio of Pd/Pt. The prepared 3 wt.% Pd-Pt (2:1)/C reached high selectivity of furfuryl alcohol (74%) with 95% conversion of furfural at 140 °C in 3 h. The catalytic activity of 3 wt.% Pd/C and 3 wt.% Pt/C catalysts had no significant decrease after recycle used for four times, confirming stable catalytic performance (Fig. 10c-d). The proposed catalyst preparation strategy avoided the use of organic solvent and catalyst deactivation, facilitated the reactants diffusion, and enhanced the biomass utilization.

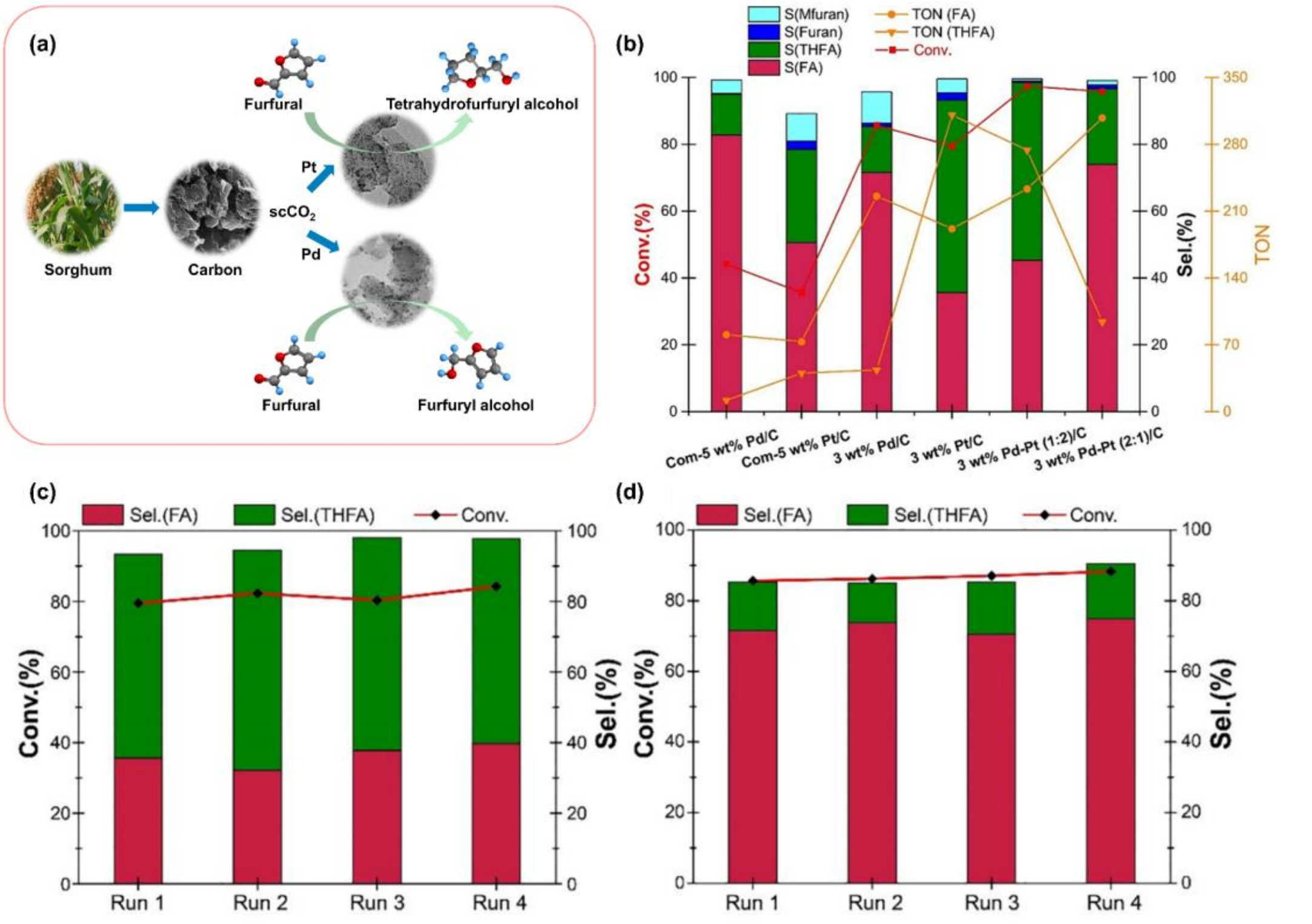

**Fig. 10.** (a) The products of furfural hydrogenation over the prepared catalysts. (b) The catalytic activity of different catalysts for furfural hydrogenation. (c) Recycle test of 3 wt.% Pt/C and (d) 3 wt.% Pd/C (Com-5wt% Pd/C: commercial 5wt% Pd/C, Com-5wt%

Pt/C: commercial 5wt% Pt/C, Mfuran: 2-methylfuran, FA: furfuryl alcohol, THFA: tetrahydrofurfuryl alcohol). Reproduced from Ref. [88] with permission from American Chemical Society, Copyright 2019.

The hydrogenation of other biomass-based aldehydes was also investigated by using $scCO_2$-assisted preparation catalysts. Ikushima et al. [118] Prepared Au nanoparticles supported on MCM-48 silica catalyst (Au-MCM-48) using the $scCO_2$ deposition method for the hydrogenation of crotonaldehyde under the reaction medium of $scCO_2$. The pressure of $scCO_2$ used in the deposition stage played an essential role on the particle size of Au on the pore of MCM-48. The particle size of Au-MCM-48 prepared using 17 MPa $CO_2$ (~2 nm) was smaller than that of prepared using 10 MPa $CO_2$ (10 nm). However, the larger Au particles were beneficial for crotonaldehyde hydrogenation to crotyl alcohol. In addition, the influence of reaction solvent on the catalytic performance was also studied. Au-MCM-48 used in the reaction media of $scCO_2$ showed the highest yield of crotyl alcohol when compared with that of reaction in propanol, acetone, and hexane. Similarly, Inomata et al. [115] prepared rhodium particles anchored on two kinds of silica (MCM-41 and MSU-H) through $scCO_2$ deposition for catalyzing the hydrogenation of cinnamaldehyde (Fig. 11a). The effects of the pore size of supports and the temperature of catalyst calcination on the catalytic performance were investigated. For MCM-41 with 2.7 nm of pore size, the high calcination temperature was conducive to the formation of large Rh nanoparticles. On the contrary, uniform Rh particles with the size of 8.4 nm on MSU-H were obtained. It indicated that the Rh/MCM-41 catalyst owned the good thermal stability and no aggregation of Rh particles under high calcination temperature. The influence of reaction solvent used in the cinnamaldehyde hydrogenation on the catalytic behavior of Rh/MCM-41 was investigated (Fig. 11b). The catalytic activity for cinnamaldehyde hydrogenation in $scCO_2$ was slightly higher than in hexane and ethanol. In addition, Rh/MSU-H displayed a better catalytic performance for the conversion of cinnamaldehyde than Rh/MCM-41 (Fig. 11c). Rh/MSU-H with a calcination temperature of 400 °C reached 47.1% conversion and 85.1% selectivity of hydrocinnamaldehyde at 50 °C in 3 h. The easy diffusion of reactants in the large pore of Rh/MSU-H may led to better activity than Rh/MCM-41.

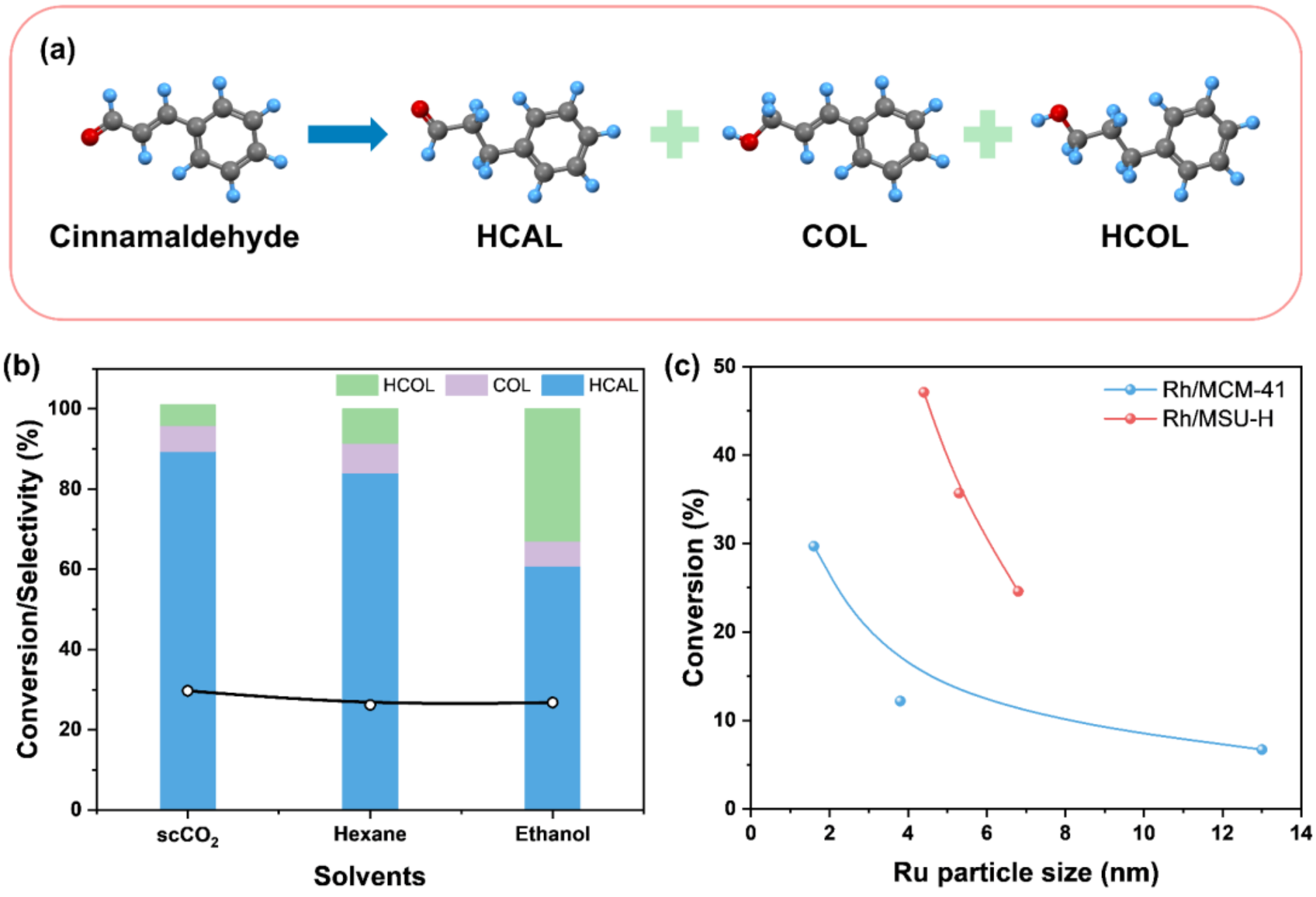

**Fig. 11.** (a) The reaction pathway of cinnamaldehyde hydrogenation. Influence of solvents (b) and Rh particle size (c) on the catalytic activity for cinnamaldehyde hydrogenation (HCAL: hydrocinnamaldehyde, COL: cinnamyl alcohol, HCOL: hydrocinnamyl alcohol). Reproduced from Ref. [115] with permission from Elsevier,

Copyright 2015.

Besides, Li et al. [119] reported the D-glucose hydrogenation into sorbitol by using carbon nanotubes (CNTs) anchored Ru catalysts via $scCO_2$ impregnation followed by the reduction in $H_2$ flow. The effects of carbon support and the medium and time of impregnation on the catalytic activity of D-glucose hydrogenation were investigated, as shown in Fig. 12. Ru/CNT catalysts all showed better catalytic performance than the Ru/AC catalyst (Fig. 12b). CNTs owned abundant delocalized electrons for adsorb the Ru metal particles resulting in the superior catalytic activity of CNTs-based catalysts. Although the initial catalytic activity of the Ru/CNT catalyst impregnated in $H_2O$ showed no noticeable difference with the Ru/CNT catalyst prepared by $scCO_2$ impregnation, the catalytic stability of the latter was better than the former. 21% reduction of catalytic activity appeared on Ru/CNT impregnated in $H_2O$ after the reaction time of 60 h (Fig. 12c). The time of deposition was also important for the D-glucose hydrogenation (Fig. 12d). The optimal time of impregnation was 1 hour for Ru/CNT catalyst, which avoided the excessive impregnation of Ru particles.

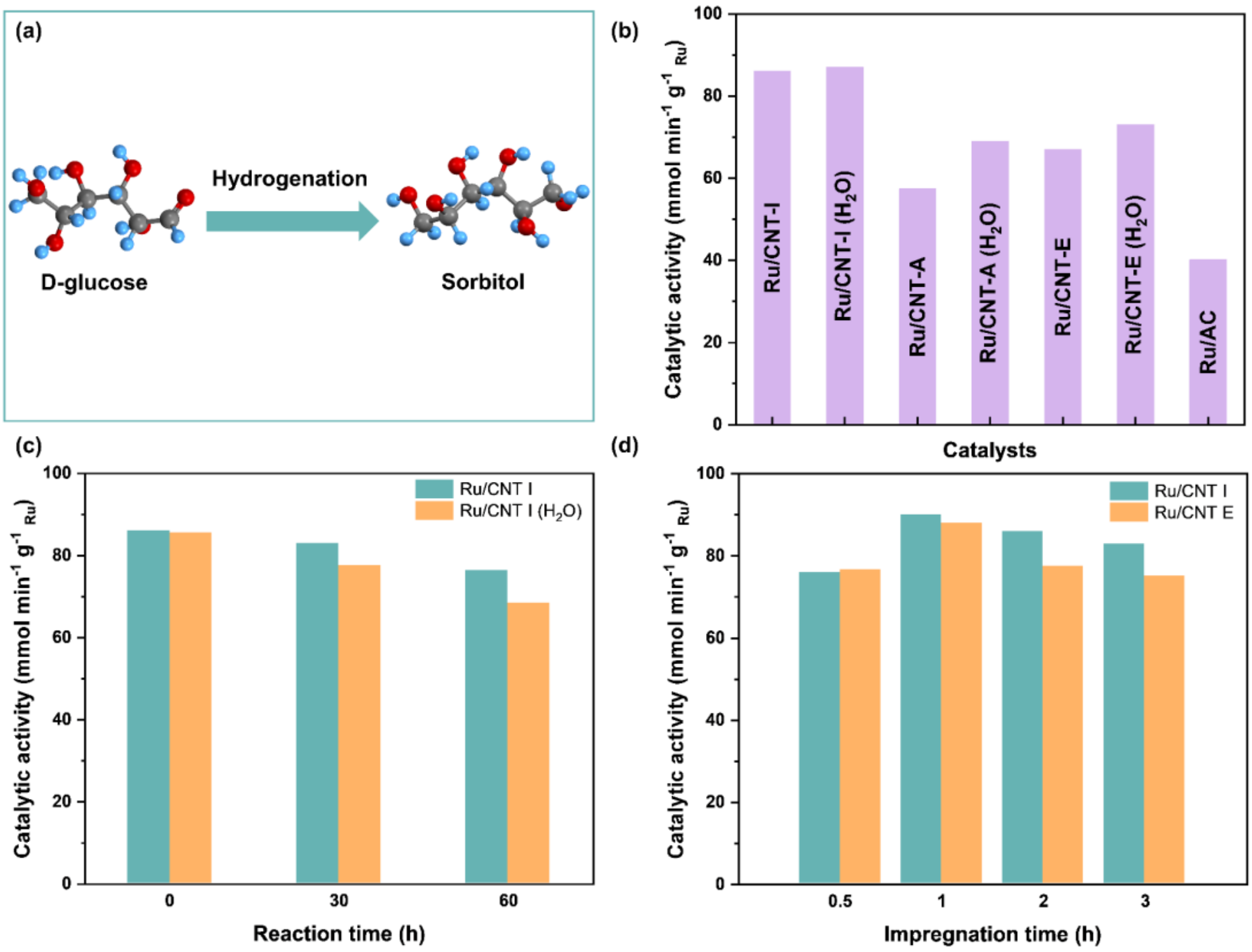

**Fig. 12.** (a) The product of the hydrogenation of D-glucose. (b) Catalytic activity of different catalysts for D-glucose hydrogenation. (c) The effect of reaction time on catalytic activity. (d) The effect of impregnation time on catalytic activity (CNT-I, CNT-A, and CNT-E: types of CNT). Reproduced from Ref. [119] with permission from Royal Society of Chemistry, Copyright 2014.

The hydrogenation of benzene rings on biomass-based aromatic compounds is an efficient route to produce value-added chemicals. Han et al. [116] prepared a Zr-based metal-organic framework supported Ru particles catalyst (Ru@Zr-MOF) under $scCO_2$-methanol for the benzene hydrogenation. Methanol acted as the bifunctional additive for the dissolution and reduction of Ru precursor. Ru@Zr-MOF obtained 99.5% conversion of benzene at 40 °C in 2 MPa $H_2$, and showed excellent catalytic stability during four recycle tests. Ru@Zr-MOF exhibited better catalytic activity than Ru/Zr-MOF synthesized via a conventional impregnation method, which was due to the uniformly dispersed Ru nanoparticles in the former (2.3 nm). Besides, Ru@Zr-MOF also displayed excellent activity for the hydrogenation of the benzene derivatives (toluene, ethylbenzene and *tert*-butylbenzene). Tan et al. [83] prepared RhPt/SBA-15 catalyst by using $scCO_2$ deposition method followed by the reduction in $H_2$ flow for the terephthalic acid (TPA) hydrogenation in an aqueous solution. To achieve the required metal loading, the metal precursors were impregnated using tetrahydrofuran before $scCO_2$ deposition. The possible products of TPA hydrogenation over RhPt/SBA-15 catalysts were shown in Fig. 13a. The addition of Rh helped the metal particles disperse better. The radio of Rh/Pt had a significant effect on the TPA conversion (Fig. 13b). There was no TPA conversion over Pt/SBA-15, while Rh/SBA-15 showed 24.2% conversion at 80 °C for 2 h. The addition of Rh also enhanced the catalytic performance of RhPt/SBA-15 catalysts. $Rh_{70}Pt_{30}$/SBA-15 reached 100% conversion of TPA at 80 °C for 4 h. There were enhanced synergistic effect of Rh and Pt in RhPt/SBA-15 catalysts, which Pt adsorbed the reactant and Rh enhanced the hydrogenation of TPA. The recycle test result of $Rh_{70}Pt_{30}$/SBA-15 showed that the TPA conversion was higher than 80% after the five cycles of reaction (Fig. 13c).

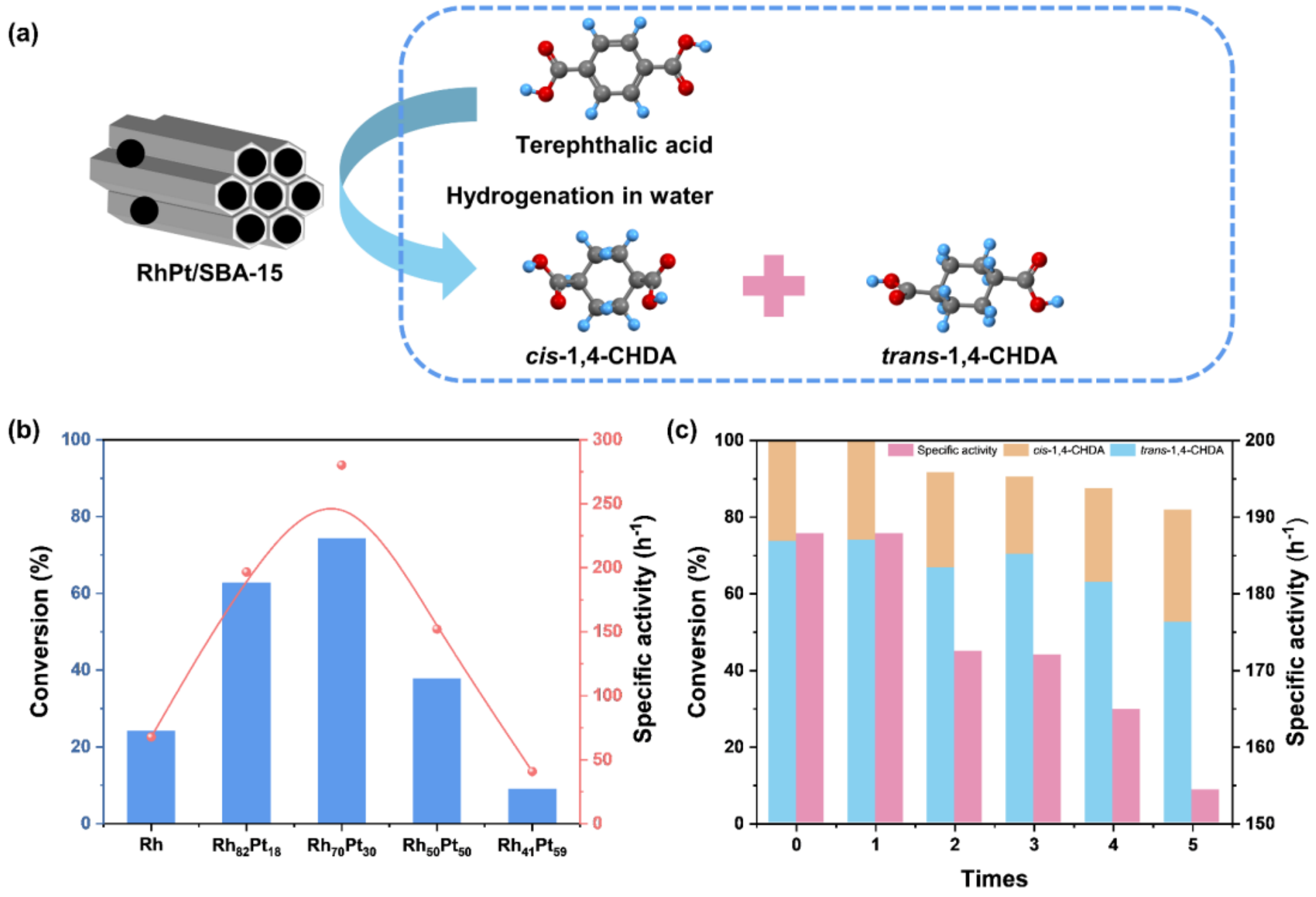

**Fig. 13.** (a) The products of TPA hydrogenation over RhPt/SBA-15 catalysts. (b) The catalytic performance of TPA hydrogenation on different catalysts. (c) Recycle tests of $Rh_{70}Pt_{30}$/SBA-15 for TPA hydrogenation under the conditions of 5 MPa $H_2$, 80 °C, and 4 h (1,4-CHDA: 1,4-cyclohexanedicarboxylic acid). Reproduced from Ref. [83] with permission from Elsevier, Copyright 2016.

$CO_2$-assisted preparation catalysts have also been used for the hydrogenation of biomass-based esters. Yin et al. [117] investigated the catalytic activity of CuO/SBA-15 catalysts prepared in sc$CO_2$ for dimethyl oxalate hydrogenation. The effect of the depressurization rate of sc$CO_2$ after the deposition stage on the particle size of CuO for the prepared CuO/SBA-15 catalysts was analyzed. Ethylene glycol and $H_2O$ were used as co-solvents to assist in the deposition. The high depressurization rate could remove the co-solvent quickly and provide a high mechanical perturbation and nucleation rate, resulting in the relatively high loading of Cu and uniform nanoparticles after calcination.

On the contrary, the low depressurization rate led to a long time to decompress and caused the large CuO nanoparticles after the subsequent calcination. The prepared CuO/SBA-15 with a high depressurization rate of 1.2 MPa/min showed high catalytic stability, reaching 100% conversion and 90% selectivity of ethylene glycol during 120 h at 200 °C under 1.2 $h^{-1}$ of WHSV.

*4.2. Oxidation*

Oxidation is another important chemical reaction for the upgrading of biomass, especially the oxidation of alcohols. Various value-added chemicals can be generated through biomass-derived alcohol oxidation. Benzyl alcohol (BAL) oxidation is one of the most essential transformations in biomass conversion. The high activity of Cu-based catalysts for BAL oxidation has been widely reported. Cheng et al. [120] prepared a novel Cu-based catalyst ($Cu_2O/Cu_3(OH)_2(CO_3)_2$) by $scCO_2$ used for the BAL oxidation. The hierarchically porous structure was not only beneficial for the exposure of active sites but also enhanced the adsorption of alcohol and mass transfer during the oxidation reaction. Furthermore, the synergistic catalytic effect of $Cu_2O$ and $Cu_3(OH)_2(CO_3)_2$ promoted the conversion of BAL. $Cu_2O/Cu_3(OH)_2(CO_3)_2$ achieved excellent catalytic performance for the BAL oxidation into benzaldehyde at 75 °C in atmospheric $O_2$. Xu et al. [112] prepared N-doped carbon-supported Cu single atoms catalysts (Cu-NC) via $scCO_2$ deposition. $Cu(acac)_2$ avoided the agglomerations of Cu atoms by providing the steric hindrance effect. The rich pore structure and N-doped sites, and large surface area favored the anchoring and distribution of Cu atoms. Besides, the physical and chemical properties of $scCO_2$ could inhibit the excessive migrations of Cu atoms. Based on these advantages, the prepared Cu-NC showed the uniform dispersion of Cu atoms on the support. The state changes for the $Cu(acac)_2$ dissolution during the deposition are shown in Fig. 14(a1-a3). There was no $Cu(acac)_2$ residue at the final state of the deposition when N-doped carbon was added. On the contrary, $Cu(acac)_2$ still existed in the cavity after deposition without NC support (Fig. 14a4). Cu-NC catalyst was used to oxide BAL to benzaldehyde by using tert-butyl hydroperoxide (TBHP) as the oxygen resource (Fig. 14b). Cu-NC reached 73.9% conversion and 63.3% yield of benzaldehyde at 50 °C for 1 h. Besides, the catalytic activity of Cu-NC showed no significant difference after four cycles (Fig. 14c). The DFT result indicated that the configuration of $CuN_3C_1$ was essential for the activation of TBHP and the production of active oxygen species (O*) (Fig. 15).

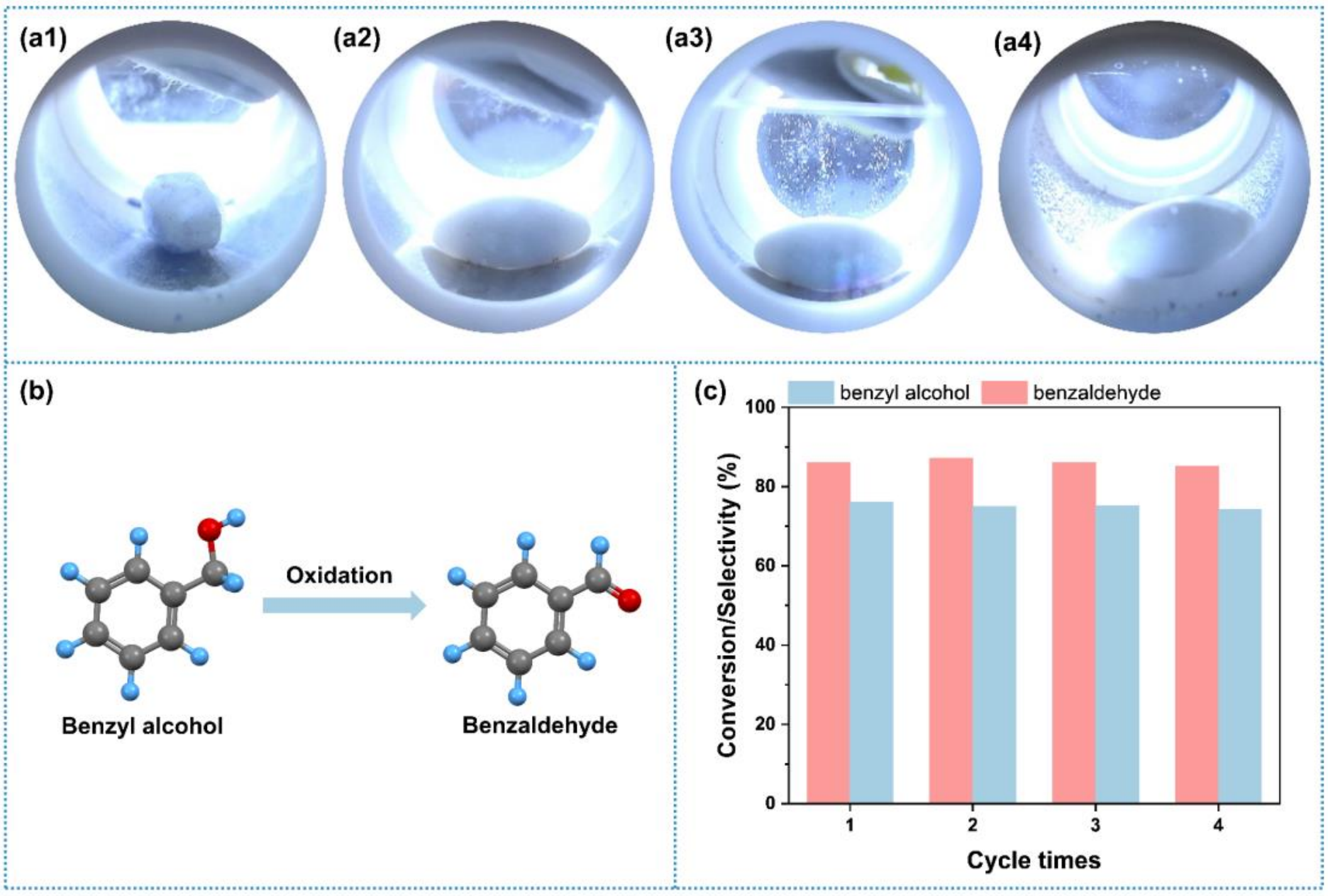

**Fig. 14.** (a1) The state of $Cu(acac)_2$ and NC supports in the $scCO_2$ (50 °C, 20 MPa) at the initial time. (a2) The final state of the deposition. (a3) The state of the released

$scCO_2$ after the deposition. (a4) The final state of the deposition is without NC support. (b) The product of benzyl alcohol oxidation. (c) Recycle test of Cu-NC. Reproduced from Ref. [112] with permission from Elsevier, Copyright 2021.

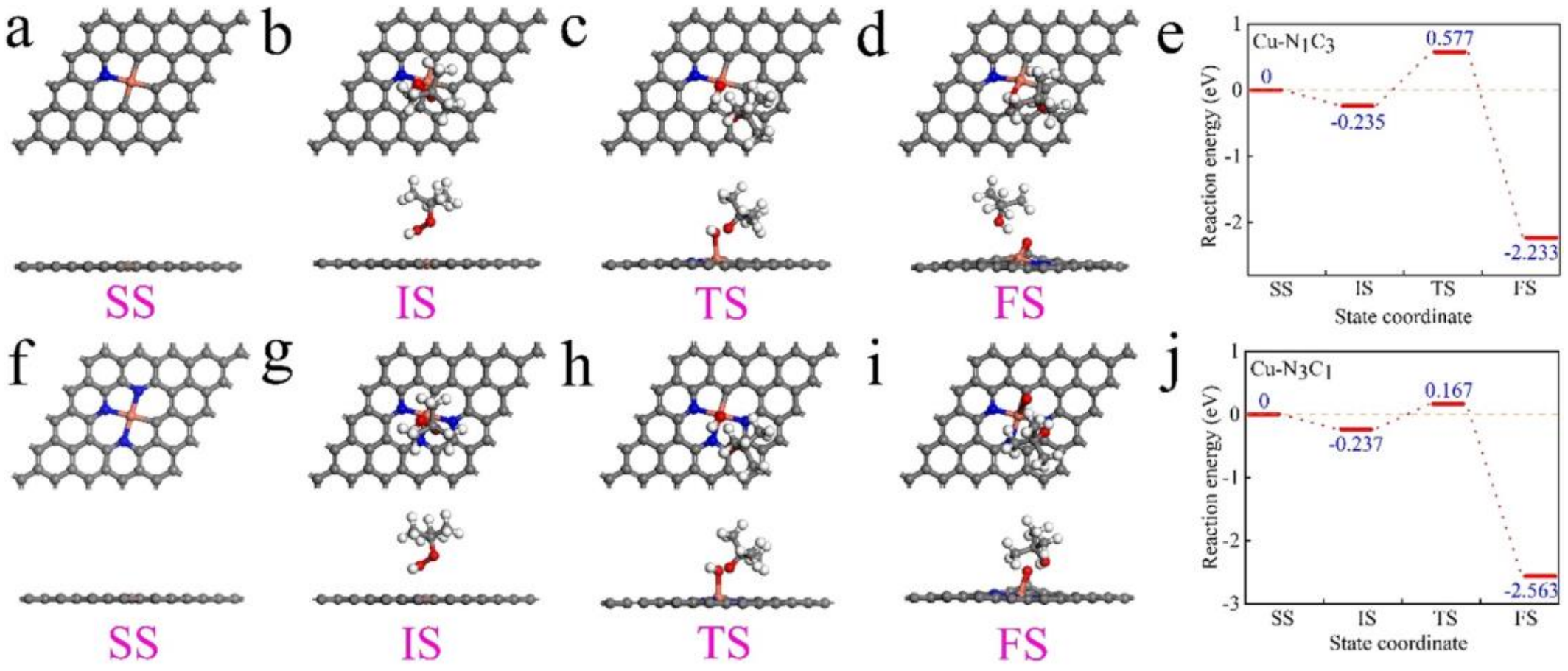

**Fig. 15.** The optimized configuration for the TBHP adsorption and dissociation of $CuN_1C_3$ (a-d) and $CuN_3C_1$ (f-i). The minimum energy pathways for the TBHP adsorption and dissociation on Cu sites of $CuN_1C_3$ (e) and $CuN_3C_1$ (j) (SS: substrate state, IS: initial structure, TS: transition structure, FS: final structure). Reproduced from Ref. [112] with permission from Elsevier, Copyright 2021.

Co-based materials were also studied as efficient catalysts for BAL oxidation. Yin et al. [121] also used N-doped carbon supported single atom catalyst through $scCO_2$ deposition without high-temperature treatment by using N, N-Dimethylformamide as a co-solvent. Besides, $CoCl_2$ was chosen as the metal precursor. The Co single atoms of the prepared catalyst are distributed on the N-doped carbon uniformly. The results indicated that the reported $scCO_2$ method was not only limited to the use of organic metal precursor but also avoided the use of harsh conditions for the preparation of the catalyst. The Co-NC single-atom catalyst was used for BAL oxidation to analyze the catalytic activity of the prepared catalyst. The yield of benzaldehyde decreased from 64.9% to 2.0% since the reaction time extended from 1 h to 6 h at 50 °C. The long reaction time led to the formation of by-products. Co-NC catalyst obtained the best catalytic activity with 77.4 % yield of benzaldehyde at 70 °C for 1 h (Fig. 16a). The catalytic activity of Co-NC had no decrease after 4 recycle used (Fig. 16b).

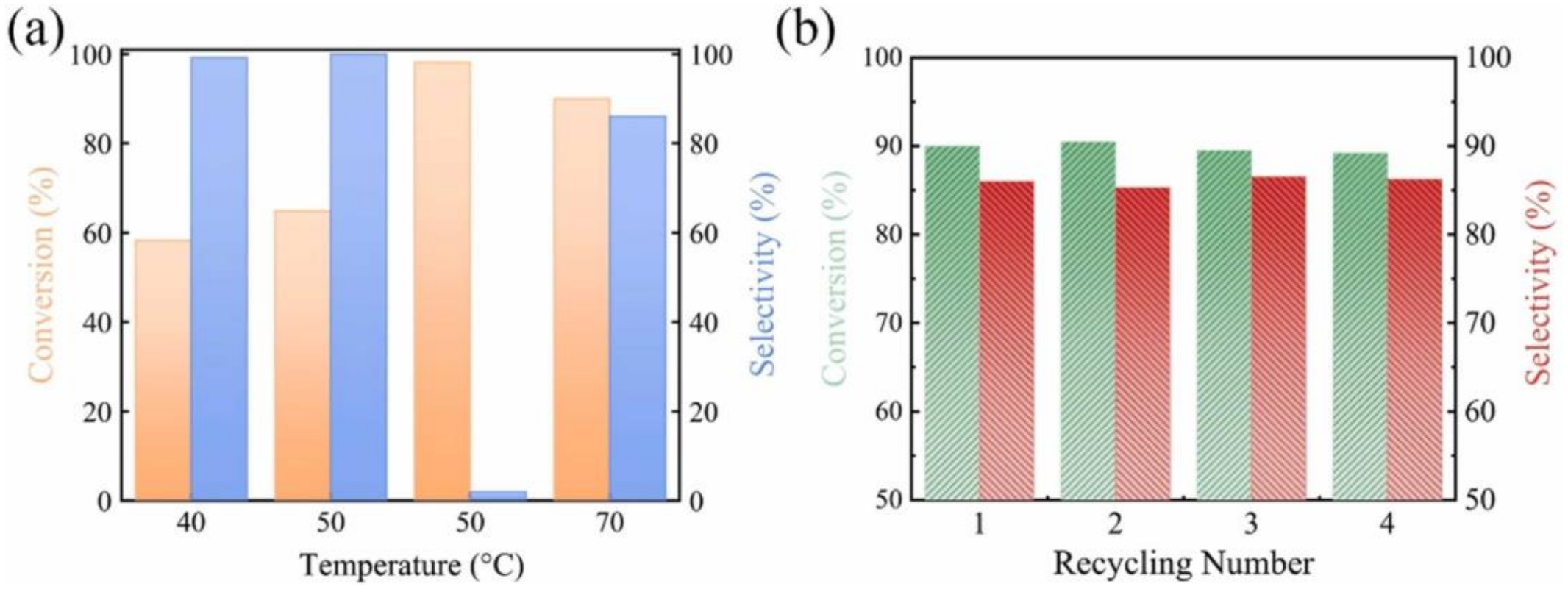

**Fig. 16.** (a) Catalytic activity of Co-NC for benzyl alcohol oxidation. (b) Recycle test of Co-NC. Reproduced from Ref. [121] with permission from Elsevier, Copyright 2022.

Yan's groups [113] also fabricated novel subnanometric amorphous $CoO_x$ clusters with high dispersion on anatase $TiO_2$ nanosheets (Co-$TiO_2$) via a $CO_2$-assisted approach. Co-$TiO_2$ could achieve the selective oxidation of various biomass-based alcohols via directly activating peroxymonosulfate (PMS). The catalytic activity for BAL oxidation over different catalysts is shown in Fig. 17a. It was obvious that Co-$TiO_2$ exhibited the highest catalytic activity. The effect of pH value on the catalytic performance was also investigated (Fig. 17b). In addition, the catalytic activity of the Co-$TiO_2$ catalyst had no noticeable changes after 4 times used (Fig. 17c). Co atoms could replace Ti atoms and adsorbed on (101) crystal plane of anatase, then acted as major active phases for BAL oxidation by activating PMS. DFT calculation and experimental results revealed that the formation of $SO_4^{\bullet-}$ and $SO_5^{\bullet-}$ was attributed to the activation of PMS by $Co^{2+}$ and $Co^{3+}$ in Co-$TiO_2$, respectively. The radicals of •OH, $SO_4^{\bullet-}$, and $SO_5^{\bullet-}$ could be generated from the redox cycle between $Co^{2+}$ and $Co^{3+}$ in Co-$TiO_2$ via transferring electrons to PMS. Subsequently, the formed •OH, $SO_4^{\bullet-}$ and the singlet oxygen ($^1O_2$) derived from $SO_5^{\bullet-}$ radicals attacked BAL to produce benzaldehyde (Fig. 17d).

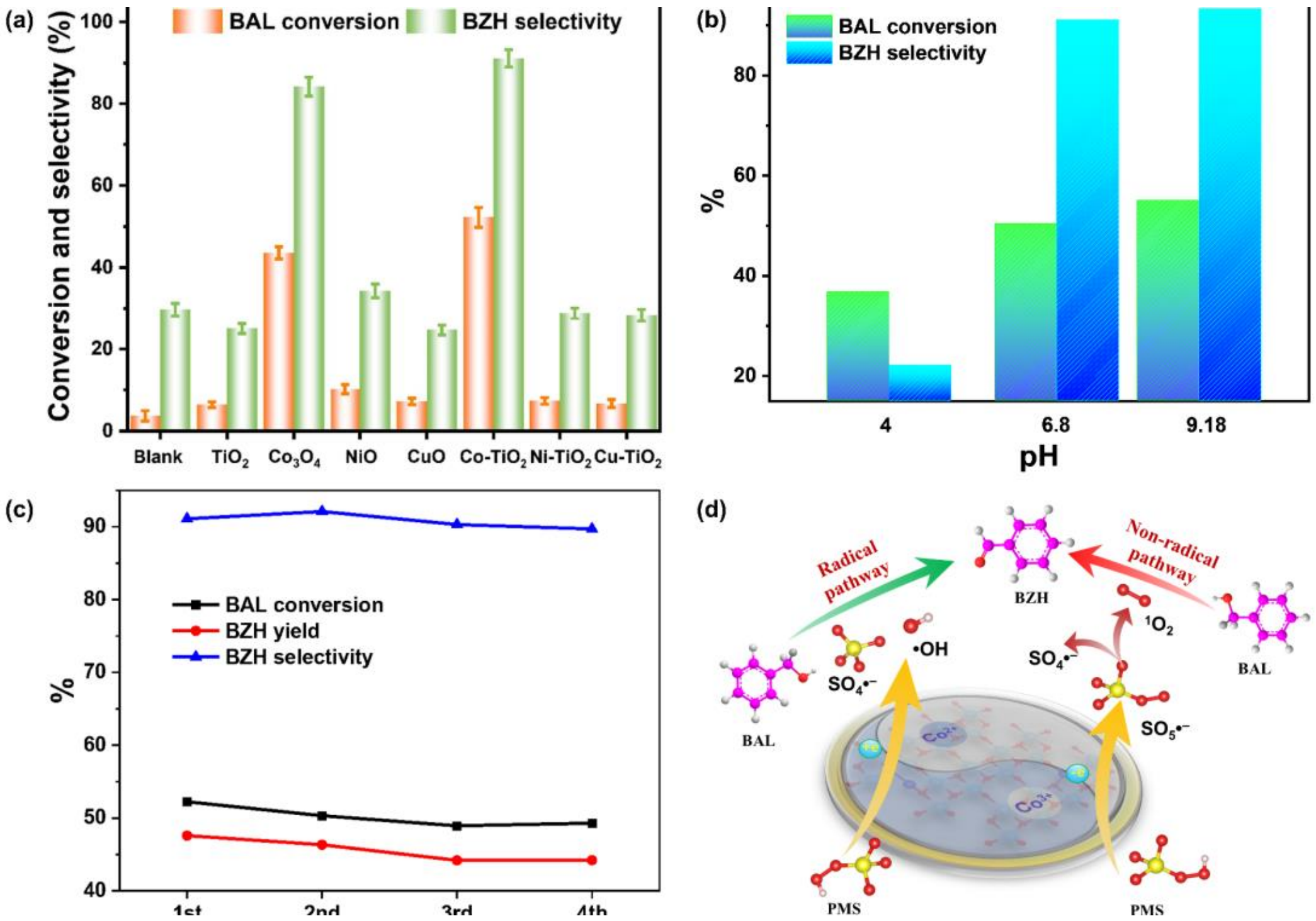

**Fig. 17.** (a) BAL oxidation over different catalysts. (b) The effect of pH on the catalytic performance. (c) Reusability of Co-$TiO_2$. (d) Reaction pathway of BAL oxidation over Co-$TiO_2$ via the activation of PMS (BZH: benzaldehyde). Reproduced from Ref. [113] with permission from Elsevier, Copyright 2023.

*4.3. Other reactions*

In addition to the hydrogenation and oxidation, the catalysts prepared by $scCO_2$ deposition are also used for the other reactions. Aymonier et al. [122] designed a series of catalysts via $scCO_2$-assisted deposition for the N-alkylation formation of aniline with benzyl alcohol. The kinds of metal and support could affect the architecture of the prepared catalysts. The surfactant used in the process of catalyst preparation was an important factor affecting the particle size and composition of the catalysts. The results of the catalytic activity test indicated that the reaction yield could be influenced by the particle size, while the surfactant and the kinds of support affected the product distribution. The prepared $CeO_2$@Pd (ethanol as cosolvent, $H_2$ as reducing source, hexadecylamine as surfactant) achieved 99% reaction yield and 68% selectivity of N-benzylaniline at 150 °C for 24 h. Zhang et al. [69] prepared Pt/$Al_2O_3$ using sc$CO_2$ deposition under the conditions of 40 °C, 110 bar and 3 h for the dehydrogenation of cyclohexane into benzene. The platinum nanoparticles on the Pt/$Al_2O_3$ catalyst synthesized via sc$CO_2$ deposition exhibited higher dispersion and showed higher catalytic performance for cyclohexane dehydrogenation into benzene than the conventional Pt/$Al_2O_3$ prepared using the impregnation method (Fig. 18a). As shown in Fig. 18b, Pt/$Al_2O_3$ catalyst synthesized via sc$CO_2$ deposition exhibited the lower apparent activation energy for cyclohexane dehydrogenation, indicating that the sc$CO_2$ deposition method effectively promoted cyclohexane dehydrogenation on Pt/$Al_2O_3$.

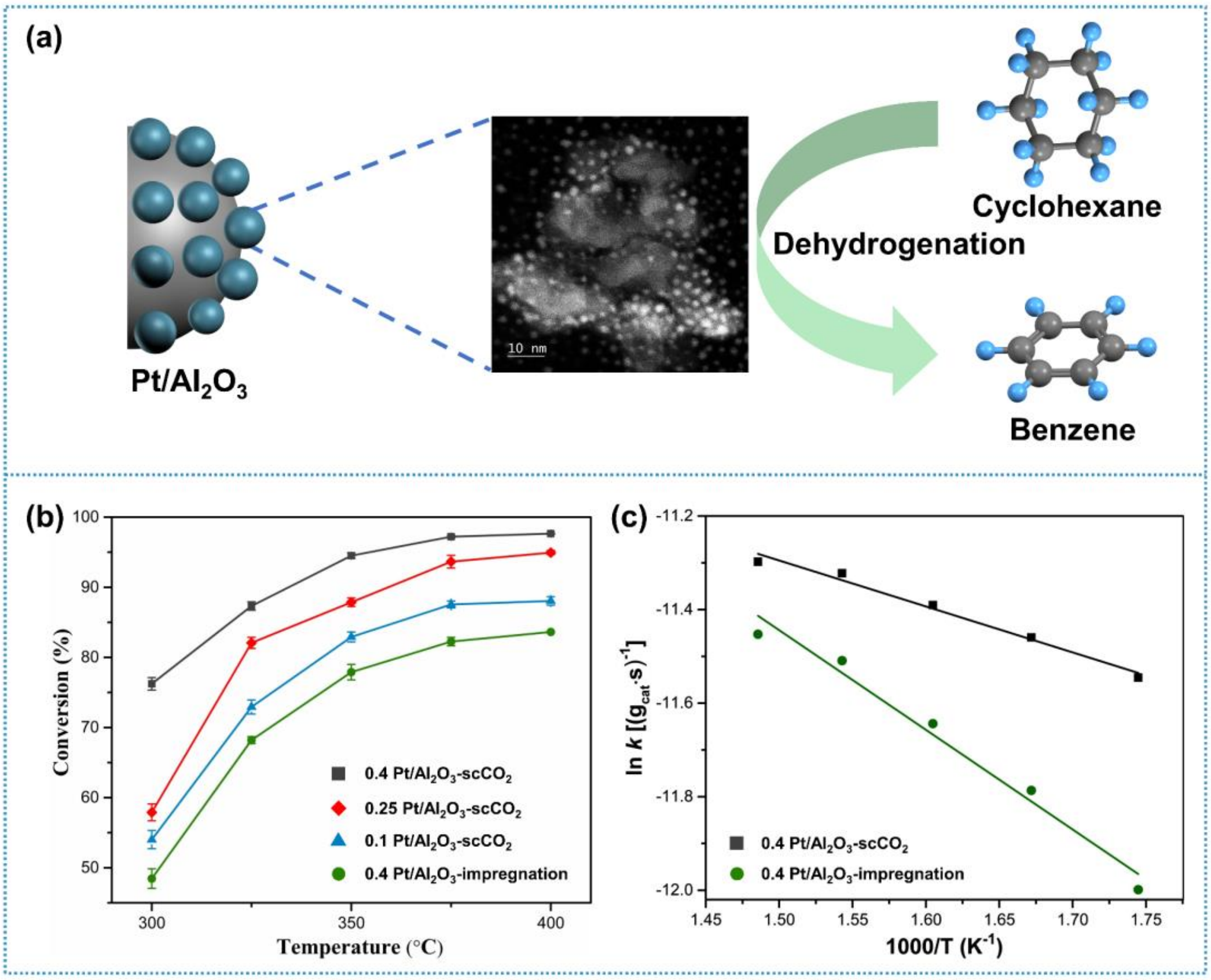

**Fig. 18.** (a) The product of the dehydrogenation of cyclohexane over $Pt/Al_2O_3$. (b) The catalytic activity of the prepared catalysts. (c) The Arrhenius plots for cyclohexane dehydrogenation of $Pt/Al_2O_3$ prepared using $scCO_2$ deposition and $Pt/Al_2O_3$ prepared using impregnation method. Reproduced from Ref. [69] with permission from Elsevier, Copyright 2020.

Irmak et al. [123] prepared a series of Pt-based supported catalysts via $scCO_2$ deposition for the gasification of wheat straw hydrolysate (Fig. 19a). The properties of the used supports had a significant effect on the sizes of Pt nanoparticles, thus affecting the catalytic activity. Carbon-based supports were beneficial to improve the dispersion of Pt nanoparticles. Among the prepared catalysts, activated carbon-supported catalyst (Pt-AC) exhibited the highest gas volume (~99 mL) and $H_2$ content (~49 mol%) from the gasification of wheat straw hydrolysate (Fig. 19a). Quitain et al. [124] prepared a multiwalled carbon nanotubes (MWCNTs) functionalized by sulfonated organosilane ($SO_3H$-MWCNTs) under $scCO_2$ used in the esterification-transesterification of kenaf oil to biodiesel. The introduction of the sulfonic group on the MWCNTs support provided sufficient acid sites, which promoted the production of biodiesel. Besides, the $scCO_2$ deposition condition facilitated organosilane compound diffusion, thus preventing the aggregation. The conversion of kenaf oil for $SO_3H$-MWCNTs prepared in $scCO_2$ (91.08%) was higher than that of the catalyst prepared via liquid chemical deposition (45.11%) (Fig. 19b).

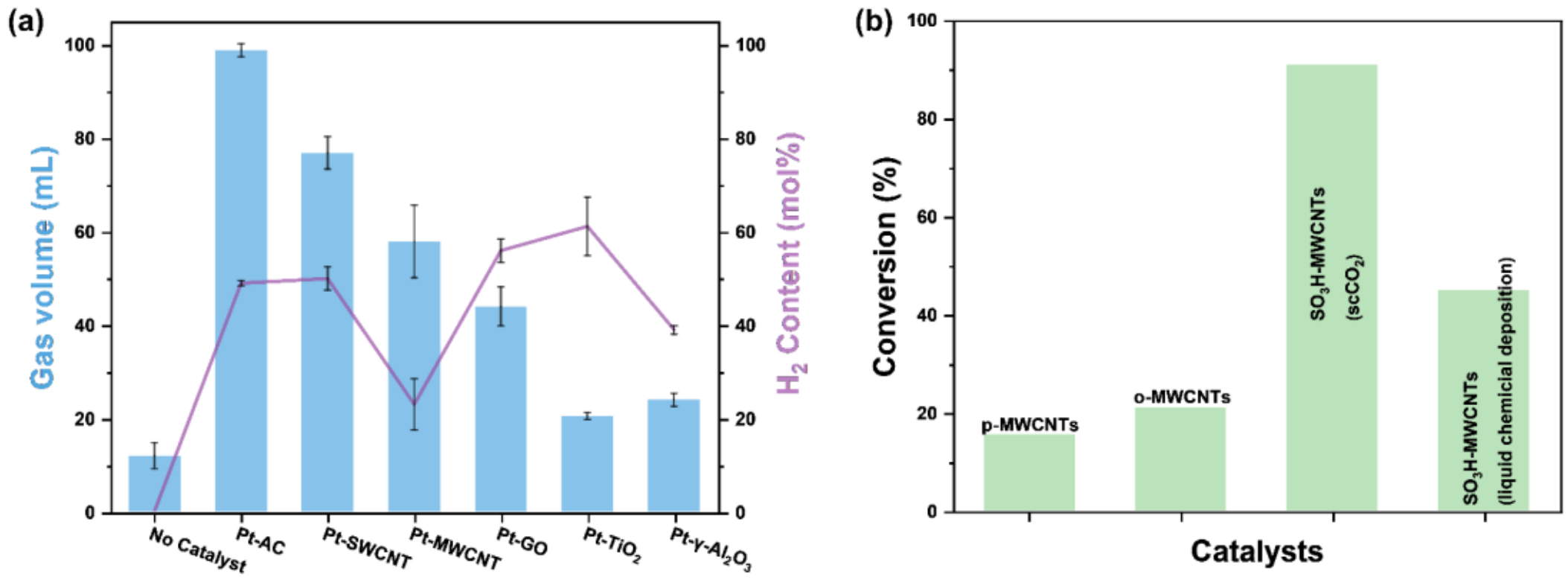

**Fig. 19.** (a) The catalytic activity for the gasification of wheat straw hydrolysate (SWCNT: single-walled carbon nanotubes). Reproduced from Ref. [123] with permission from Elsevier, Copyright 2012. (b) The catalytic performance for the conversion of kenaf oil to biodiesel (p-MWCNTs: pristine MWCNT, o-MWCNTs: oxidized MWCNT). Reproduced from Ref. [124] with permission from Elsevier, Copyright 2020.

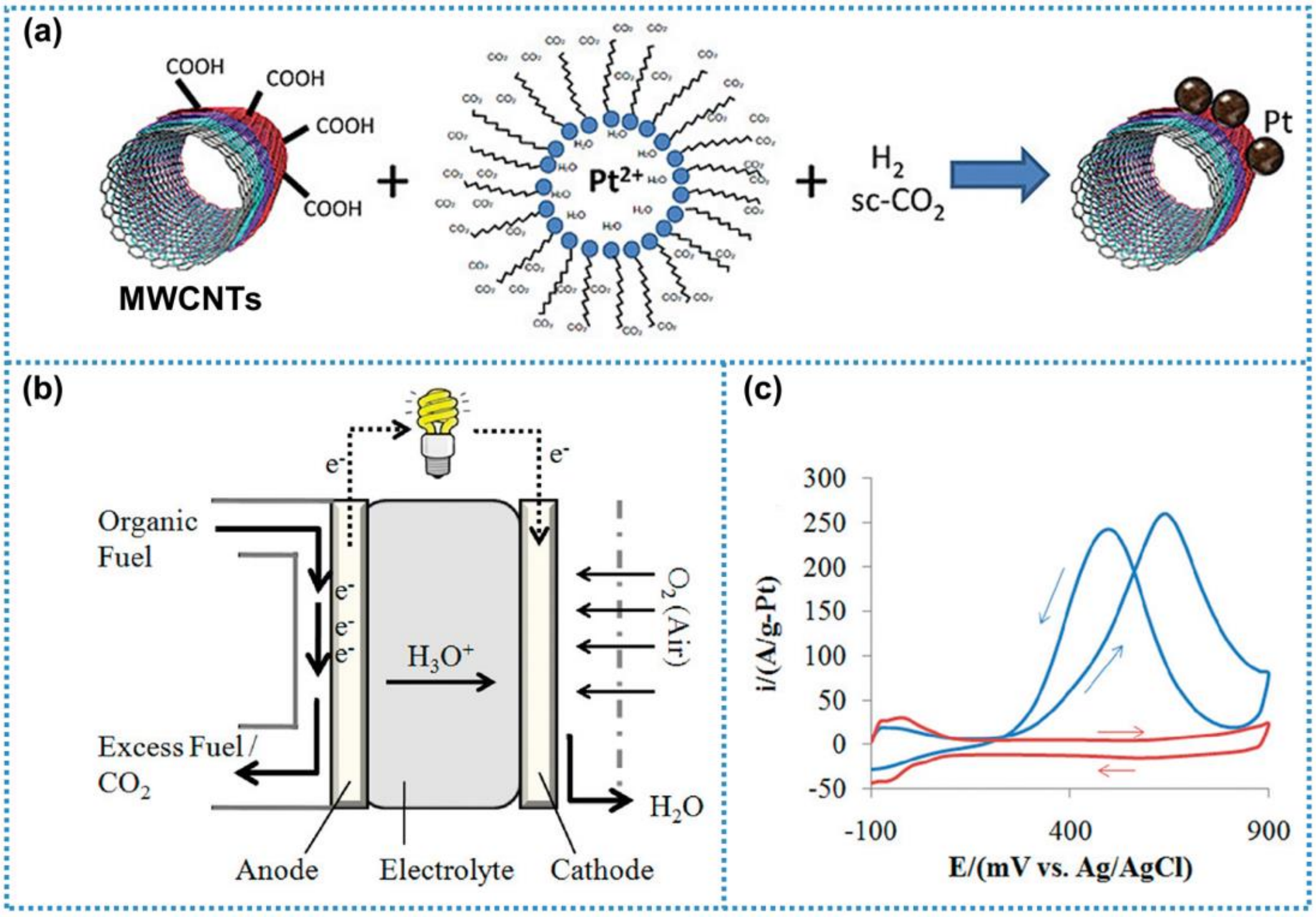

**Fig. 20.** (a) Pt deposited on MWCNTs via water-in-sc$CO_2$ microemulsion and hydrogen. (b) Scheme of direct methanol fuel cell. (c) CVs of methanol oxidation on Pt/MWCNTs in 0.1 M methanol with 1 M $H_2SO_4$ at 50 mV/s. Reproduced from Ref. [125] with permission from American Chemical Society, Copyright 2010.

Methanol and formic acid are the most common organic fuels, which can be oxidized by catalysts and then used for direct organic fuel cells. Some research focused on the preparation of catalysts via sc$CO_2$ method applied in fuel cells [81,126,127]. Wai et al. [125] firstly mixed the $Na_2PtCl_4$ solution, sodium bis(2-ethylhexyl)sulfosuccinate solution, hexane, and pretreated MWCNTs to form the water-in-hexane microemulsion, then $H_2$ and sc$CO_2$ was introduced to prepare Pt/MWCNTs catalyst (Fig. 20a). The prepared Pt/MWCNTs catalyst was used for the direct methanol fuel cell (Fig. 20b). The cyclic voltammograms (CVs) of methanol oxidation on Pt/MWCNTs (Fig. 20c) indicated the high catalytic performance of the catalyst for methanol oxidation. In fact, the conventional surfactant was easy to remain on the catalyst surface, thus deactivating the catalyst. The $scCO_2$ can produce more dispersed Pt nanoparticles and remove the residue surfactant effectively, showing its prospect for the preparation of methanol fuel cell catalysts. Zhao et al. [76] also found that Pd/pristine graphene (Pd/PG) synthesized by $scCO_2$ strategy was active for the electro-oxidation of formic acid and methanol. Worm-like Pd nanoparticles deposited on the PG surfaces under the assistance of $scCO_2$ (Fig. 21a). The support effect was investigated by comparing the CV results of different supported catalysts (Fig. 21 b-c). Pd/PG showed the highest mass activity for the oxidation of formic acid (1045.7 mA $mg^{-1}$ Pd) and methanol (823.4 mA $mg^{-1}$ Pd) when compared with other carbon materials supported catalysts. In addition, the charge transfer resistance value of Pd/PG (101.4 Ω) was also the lowest in comparison with the prepared catalysts in 0.5 M HCOOH with 0.5 M $H_2SO_4$ at 0.2 V (vs Ag/AgCl). The graphene structure not only improved the catalytic stability of Pd/PG, but also facilitated the electron transport, which enhanced the catalytic activity of Pd/PG.

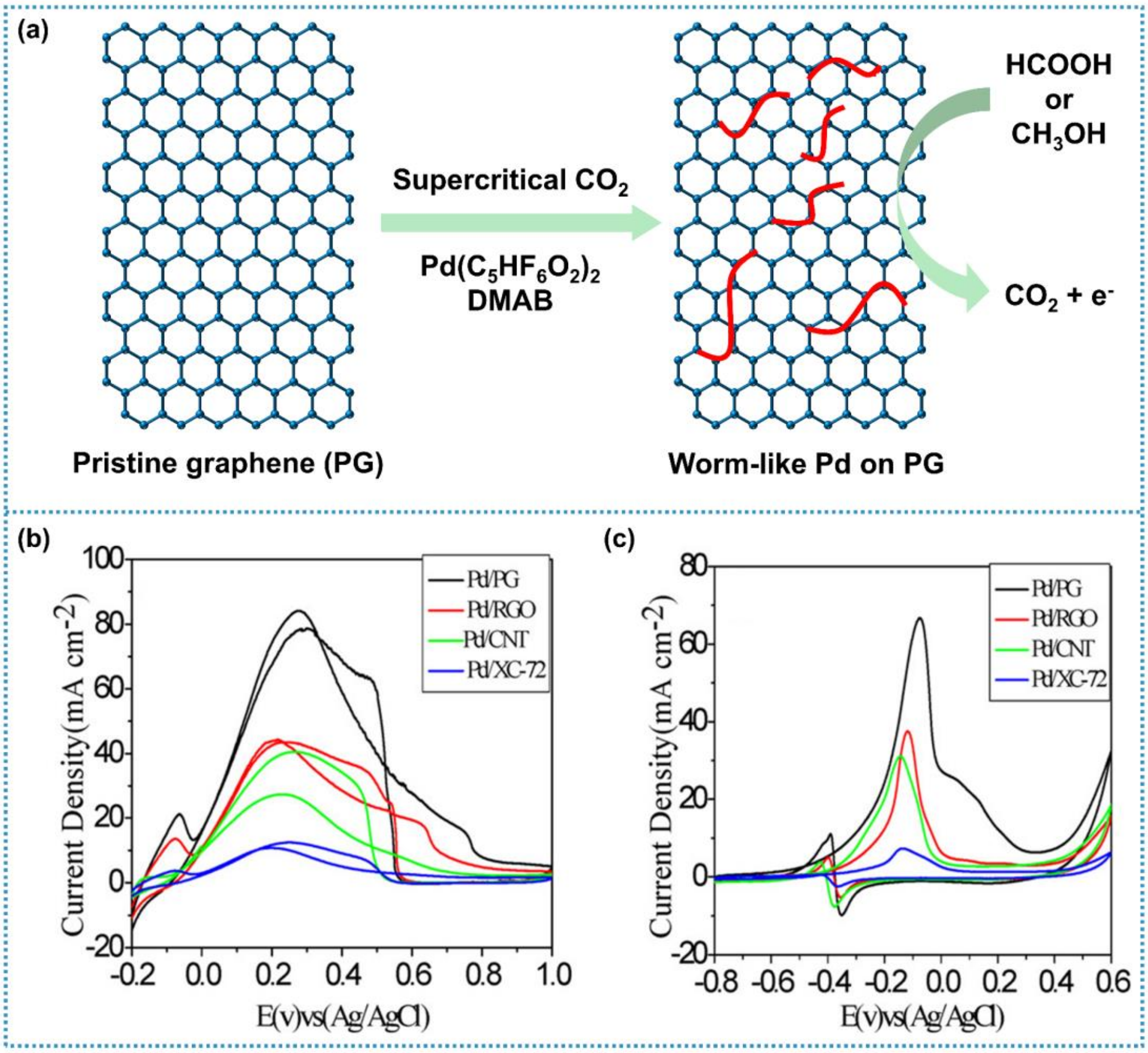

**Fig. 21.** (a) Scheme of Pd deposited on pristine graphene via sc$CO_2$ for fuel cell. (b) CVs of formic acid oxidation in 0.5 M formic acid with 0.5 M $H_2SO_4$ and (c) CVs of methanol oxidation in 1 M methanol with 0.5 M NaOH on the prepared catalysts at 50 mV/s. Reproduced from Ref. [76] with permission from American Chemical Society, Copyright 2015.

## 5. Conclusions and outlook

This review demonstrated the potential of sc$CO_2$ for the generation of metal nanoparticles, oxides, and other materials. The nanoparticles' morphological and physicochemical characteristics can be changed depending on the interaction between metal complexes and the surface of the support material. Pd, Pt, Ru, Ni monometallic and bimetallic nanoparticles have been synthesized through $CO_2$-assisted. The effect on of support, reaction temperature, the metal concentration and the $CO_2$ pressure on the size of metal nanoparticle were systematically investigated. Some different metal oxide nanoparticles and single metal atom materials were also simply summarized. Subsequently, these as-prepared metal nanoparticles were used as catalysts for the conversion of biomass-derived chemicals. The hydrogenation, oxidation, animation, dehydrogenation and other reactions of biomass-derived chemicals are mainly discussed in this review. The effects of supports, metal species, metal concentration, and metal component of catalysts have been examined for the conversion reactions.

Although significant progress has been achieved for the upgrading of biomass-derived compounds using the metal nanoparticles through $CO_2$-assisted preparation, some issues are still existed: 1) Development novel methods for metal-based catalysts through $CO_2$ SCFs with cost-effective high-performance is highly desired; 2) Controllable preparation of uniform metal nanoparticles is necessary; 3) the formation mechanisms of metal nanoparticles are limited and need further study by in-situ TEM; 4) Computational simulation, along with experimental methods, could be an alternative to describe through modeling the properties of nanoparticles, such as selectivity and reaction conversion, according to the $CO_2$ SCFs phase; 5) Exploring the synergy between process variables (flow rate, contact time, temperature, and pressure) and economic feasibility studies are relevant to promoting this fashionable ecological technology; 6) more reaction styles for biomass conversion using metal nanoparticle through SCFs need to be developed; 7) Reaction mechanism and process (C-C/O bond cleavage) should be further investigated by DFT and in-situ tools such as in-situ FTIR, in-situ Raman. The present review provides a new route for the synthesis of high-performance metal nanoparticles using green $CO_2$ SCFs and discusses their applications in biomass conversion into high-value chemicals. This review could be helpful for the rational design of more efficient metal catalysts for selective synthesis of fine chemicals and fuels from biomass-derived chemicals.

**Declaration of competing interest**

The authors declare that they have no known competing financial interests or personal relationships that could have appeared to influence the work reported in this paper.

**Data availability**

Data will be made available on request.

**Acknowledgments**

This work was supported by National Key R&D Program of China (2023YFC3905804), the National Natural Science Foundation of China (22078374, 22378434), Key Realm Research and Development Program of Guangdong Province (2020B0202080001), the Guangdong Basic and Applied Basic Research Foundation (2019B1515120058), Science and Technology Planning Project of Guangdong Province, China (2021B1212040008), Guangdong Laboratory for Lingnan Modern Agriculture Project (NT2021010), the Scientific and Technological Planning Project of Guangzhou (202206010145).